\pdfoutput=1 
\documentclass[draft, table]{agujournal2019}
\usepackage{xcolor} 
\usepackage{url} 
\usepackage{lineno}
\usepackage[inline]{trackchanges} 
\usepackage{soul}
\usepackage{amssymb}
\usepackage{gensymb}
\usepackage{textcomp, array, multirow, bigdelim, makecell, rotating} 
\usepackage{pifont} 
\usepackage{booktabs} 
\usepackage{xr} 

\draftfalse

\journalname{Journal of Advances in Modeling Earth Systems (JAMES)}

\makeatletter
\@ifundefined{c@rownum}{%
  \let\c@rownum\rownum
}{}
\@ifundefined{therownum}{%
  \def\therownum{\@arabic\rownum}%
}{}
\makeatother

\newcommand*\CHECK{\ding{51}}

\makeatletter
\newcommand*{\addFileDependency}[1]{
  \typeout{(#1)}
  \@addtofilelist{#1}
  \IfFileExists{#1}{}{\typeout{No file #1.}}
}
\makeatother

\newcommand*{\myexternaldocument}[1]{%
    \externaldocument{#1}%
    \addFileDependency{#1.tex}%
    \addFileDependency{#1.aux}%
}

\myexternaldocument{supp}

\begin{document}

%
%

\graphicspath{{figures/}}

\title{Deep Learning Based Cloud Cover Parameterization for ICON}

%
%




\authors{Arthur Grundner\affil{1,2}, Tom Beucler\affil{3}, Pierre Gentine\affil{2}, Fernando Iglesias-Suarez\affil{1}, Marco A. Giorgetta\affil{4}, and Veronika Eyring\affil{1,5}}

\affiliation{1}{Deutsches Zentrum für Luft- und Raumfahrt e.V. (DLR), Institut für Physik der Atmosphäre, Oberpfaffenhofen, Germany}
\affiliation{2}{Columbia University, Center for Learning the Earth with Artificial intelligence And Physics (LEAP), New York, NY 10027, USA}
\affiliation{3}{University of Lausanne, Institute of Earth Surface Dynamics, Lausanne, Switzerland}
\affiliation{4}{Max Planck Institute for Meteorology, Hamburg, Germany}
\affiliation{5}{University of Bremen, Institute of Environmental Physics (IUP), Bremen, Germany}




\correspondingauthor{Arthur Grundner}{Arthur.Grundner@dlr.de}




\begin{keypoints}
\item Neural networks can accurately learn sub-grid scale cloud cover from realistic regional and global storm-resolving simulations 
\item Three neural network types account for different degrees of vertical locality and differentiate between cloud volume and cloud area fraction  
\item Using a game theory based library we find that the neural networks tend to learn local mappings and are able to explain model errors 

\end{keypoints}

\begin{abstract}
A promising approach to improve cloud parameterizations within climate models and thus climate projections is to use deep learning in combination with training data from storm-resolving model (SRM) simulations. The ICOsahedral Non-hydrostatic (ICON) modeling framework permits simulations ranging from numerical weather prediction to climate projections, making it an ideal target to develop neural network (NN) based parameterizations for sub-grid scale processes. Within the ICON framework, we train NN based cloud cover parameterizations with coarse-grained data based on realistic regional and global ICON SRM simulations. We set up three different types of NNs that differ in the degree of vertical locality they assume for diagnosing cloud cover from coarse-grained atmospheric state variables. The NNs accurately estimate sub-grid scale cloud cover from coarse-grained data that has similar geographical characteristics as their training data. Additionally, globally trained NNs can reproduce sub-grid scale cloud cover of the regional SRM simulation. Using the game-theory based interpretability library SHapley Additive exPlanations, we identify an overemphasis on specific humidity and cloud ice as the reason why our column-based NN cannot perfectly generalize from the global to the regional coarse-grained SRM data. The interpretability tool also helps visualize similarities and differences in feature importance between regionally and globally trained column-based NNs, and reveals a local relationship between their cloud cover predictions and the thermodynamic environment. Our results show the potential of deep learning to derive accurate yet interpretable cloud cover parameterizations from global SRMs, and suggest that neighborhood-based models may be a good compromise between accuracy and generalizability.
\end{abstract}


\section*{Plain Language Summary}
Climate models, such as the ICON climate model, operate on low-resolution grids, making it computationally feasible to use them for climate projections. However, physical processes --especially those associated with clouds-- that happen on a sub-grid scale (inside a grid box) cannot be resolved, yet they are critical for the climate. In this study, we train neural networks that return the cloudy fraction of a grid box knowing only low-resolution grid-box averaged variables (such as temperature, pressure, etc.) as the climate model sees them. We find that the neural networks can reproduce the sub-grid scale cloud fraction on data sets similar to the one they were trained on. The networks trained on global data also prove to be applicable on regional data coming from a model simulation with an entirely different setup. Since neural networks are often described as black boxes that are therefore difficult to trust, we peek inside the black box to reveal what input features the neural networks have learned to focus on and in what respect the networks differ. Overall, the neural networks prove to be accurate methods of reproducing sub-grid scale cloudiness and could improve climate model projections when implemented in a climate model.

%
%

%


%
%
%
%

\section{Introduction}

Clouds play a key role in the climate system. They regulate the hydrologic cycle and have a substantial influence on Earth's radiative budget \cite{allen2002constraints}. Yet, in climate models with horizontal resolutions commonly on the order of 100\,km, clouds are sub-grid scale phenomena, i.e., they cannot be directly resolved but need to be ``parameterized''. These parameterizations are a major cause of uncertainties in climate model projections \cite<e.g.>{randall2003, schneider2017} and effective climate sensitivity \cite{meehl2020, schlund2020emergent}.

The long-standing deficiencies in cloud parameterizations have motivated the development of high-resolution global cloud-resolving climate models \cite{klocke2017, stevens2019dyamond} with the ultimate goal of explicitly resolving clouds and convection. Yet, these simulations are extremely computationally demanding and cannot be run on climate timescales for multiple decades or for ensembles. Deep learning for the parameterization of sub-grid scale processes has been identified as a promising approach to improve parameterizations in climate models and to reduce uncertainties in climate projections \cite{eyring2021, gentine2021deep}. 

In the atmospheric component of the state-of-the-art ICOsahedral Non-hydrostatic (ICON) climate model (ICON-A), clouds result from an interplay of different parameterization schemes \cite{giorgetta2018}. In it, the cloud cover scheme takes an integral role. Its cloud cover directly influences the tendencies and hence statistics of cloud liquid water, cloud ice, and water vapor through the microphysics scheme \cite{lohmann1996, pincus2013}, and the energy balance through the radiation scheme.

Our goal is to develop a machine learning based parameterization that can replace ICON's semi-empirical cloud cover scheme. We cover the background of these two fields in section \ref{sec:background} by reviewing a) the existing cloud cover scheme in ICON (in \ref{sec:iconclc}) and b) the machine learning based parameterizations (in \ref{sec:MLparam}) before defining the scope of our study in section \ref{sec:scope}.
\subsection{Background}
\label{sec:background}
\subsubsection{Existing Cloud Cover Scheme in ICON}
\label{sec:iconclc}
 Cloud cover is estimated as a diagnostics in ICON, which is based on the local amount of relative humidity (RH), and a semi-empirical relationship devised by \citeA{sundqvist1989} and further adapted by \citeA{xu1991} (see \citeA{lohmann1996}) and \citeA{mauritsen2019}. In this scheme, cloud cover exists whenever RH exceeds a specified lower bound (the \textit{critical RH threshold}), which depends solely on atmospheric and surface pressure.

RH-based cloud cover schemes have some notable drawbacks. First of all, knowing RH does not fully determine cloud cover. For instance \citeA{walcek1994} had shown, that with an RH of 80\% and a pressure between 800 and 730\,hPa, the probability of observing any amount of cloud cover can be nearly uniform. In addition, no clear critical RH threshold seems to exist. Furthermore, even though they influence cloud characteristics, RH-based schemes do not directly differentiate between local dynamical conditions \cite<e.g., whether the grid column undergoes deep convection;>{tompkins2005parametrization}.
The ICON-A cloud cover scheme also does not account for vertical sub-grid scale cloud cover variability. An exception to this is the recent adaptation to artificially increase RH in regions below subsidence inversions to incorporate thin marine stratocumuli \cite{mauritsen2019}. \\
Finally, most cloud schemes are based on local thermodynamic variables, yet rapid advection (e.g., updrafts) could lead to non-locality in the relationship. Overall, the formation and dissipation of clouds is still poorly understood \cite{stensrud2009}. Therefore, physics-based cloud parameterizations have to build on incomplete knowledge and are prone to inaccuracies. They usually also contain tuning parameters. In the ICON-A cloud cover scheme these are the RH for 100\,\% cloud cover, the asymptotic critical RH in the upper troposphere, the critical RH at the surface, and the shape factor. These parameters have to be adjusted following the primary goal of a well balanced top-of-the-atmosphere energy budget \cite{giorgetta2018}.


\subsubsection{Machine Learning Based Parameterizations}
\label{sec:MLparam}
The field of machine learning based parameterizations is growing and can loosely be classified into two groups: The first group consists of studies about machine learning based parameterizations that emulate and speed up existing parameterizations. In \citeA{beucler2020, gentine2018, han2020, griffin2020, wang2022stable} these existing parameterizations were  superparameterizations, i.e., embedded two-dimensional cloud-resolving models \cite{khairoutdinov2005}. For instance, in a pioneering study by \citeA{rasp2018pnas}, a neural network (NN) was successfully trained to estimate sub-grid scale convective effects by learning from the output of the superparameterized Community Atmosphere Model in an idealized aquaplanet setting. Other notable members of this group, that focused on emulating more traditional parameterizations, are \citeA{chevallier2000, chantry2021, gettelman2021, krasnopolsky2005, seifert2020}. The second group consists of studies about machine learning based parameterizations that learn from three-dimensional, high-resolution data. In most of those studies, the high-resolution data was coarse-grained to the low-resolution grid of the climate model. The first proof of concept was established by \citeA{krasnopolsky2013} who trained a very small NN on coarse-grained regional data.
Later, \citeA{brenowitz2018, brenowitz2019, brenowitz2020, yuval2020, yuval2021} adapted this approach. However, in contrast to our study, they worked with idealized aquaplanet simulations and coarse-graining limited to the horizontal dimension.

While some of these studies were conducted in a purely `offline' fashion, that is, decoupled from the dynamics of the climate model, \citeA{brenowitz2019, brenowitz2020, chantry2021, gettelman2021, krasnopolsky2005, ott2020, rasp2018pnas, wang2022stable, yuval2020, yuval2021} also achieved stable online simulations in specific setups. 

Recent research has suggested that emulating sub-grid scale physics on a process-by-process level may lead to more stable machine learning powered climate simulations \cite{yuval2021}. It may also facilitate interpretability and targeted studies of the interaction between large-scale (thermo)dynamics and cloudiness.

\subsection{Machine Learning Based Cloud Cover Parameterization}
\label{sec:scope}
In the context of these new advances, our study is the first machine learning based approach specifically focused on the parameterization of cloud cover.

Our novel approach to a cloud cover parameterization is based on the idea of training a supervised deep learning scheme to estimate cloud cover from the thermodynamical state, using coarse-grained high-resolution data. We allow for vertical sub-grid scale cloud cover variability by learning the fraction of a grid volume that is cloudy \cite<`cloud volume fraction';>{brooks2005}. Cloud volume fraction is the preferable measure of cloud cover, for instance in ICON's microphysics scheme where in-cloud condensation and evaporation rates are multiplied by the volume fraction of the grid box that is cloudy \cite{lohmann1996}. In section \ref{chap:global}, we also introduce NNs that predict the horizontally projected amount of cloudiness inside a grid cell (`cloud area fraction'). The reason is that we still require cloud area fraction as a parameter for the (ICON's two-stream) radiation scheme \cite{pincus2013} to evaluate whether radiation penetrates through a cloud or not. 

The ICON modeling framework is used in realistic conditions on a variety of timescales and resolutions \cite{zaengl2015}. It thus allows us to work with data from high-resolution ICON simulations to train machine learning based parameterizations fit for the low-resolution ICON climate model. Observations, on the other hand, are temporally and spatially sparse and would thus constitute less adequate training data \cite{rasp2018pnas}. The basis of our training data form new storm-resolving ICON simulations from the Next Generation Remote Sensing for Validation Studies (NARVAL) flight campaigns \cite{stevens2019} and the Quasi-Biennial Oscillation in a Changing Climate (QUBICC) project \cite{giorgetta2022}, with horizontal resolutions of 2.5\,km and 5\,km respectively. At these resolutions one can generally consider deep convection to be resolved \cite{vergara-temprado2020}, and therefore these simulations forego the use of convective parameterizations. \citeA{hohenegger2020} systematically compared 27 different statistics in ICON simulations with resolutions ranging from 2.5\,km to 80\,km. They concluded that simulations with explicit convection at resolutions of 5\,km or finer may indeed be used to simulate the climate. \citeA{stevens2020} have shown that the NARVAL simulations can more accurately represent clouds and precipitation than simulations with an active convective parameterization.

We train NNs on coarse-grained data from these high-resolution simulations. Here, two commonly used ICON-A grids (with horizontal resolutions of 80\,km and 160\,km) are the target grids we coarse-grain to. ICON uses an icosahedral grid in the horizontal and a terrain-following height grid in the vertical. On these grids, more sophisticated and partly new methods of coarse-graining are required than on simpler regular grid types. As our machine learning algorithm we choose NNs, which are able to incorporate this wealth of data to --in principle-- approximate any type of nonlinear function \cite{gentine2018, hornik1991approximation}. While being generally fast at inference time, NNs also have computational advantages over alternative machine learning based approaches such as random forests \cite{yuval2021}. Hence, an NN-powered parameterization of cloud cover could accelerate and improve the representation of cloud-scale processes (from radiative feedbacks to precipitation statistics).

In this study, we focus on developing an \textit{offline} (i.e., without coupling to the dynamical core), ML-based cloud cover parameterization for ICON. While offline skill does not always guarantee \textit{online} performance once the NN is coupled back to the dynamical core \cite{gagne2020}, \citeA{ott2020} showed that offline skill generally correlated with the stability (although not necessarily the accuracy) of online simulations. Several time-consuming tasks are required to achieve operational online skill, such as ensuring excellent extrapolation skills to different distributions of state variables for stable simulations (across climate-regimes). Then, a re-calibration of the coarse-resolution climate model against the observed state of the atmosphere (top-of-the-atmosphere radiative fluxes, global mean surface temperature, clouds, precipitation, wind fields, etc., \citeA{giorgetta2018}) is most likely necessary, for example, since there are too few (low-level) clouds in the ICON model, and other tunable parameters are currently calibrated to compensate for that fact \cite{crueger2018icon}. After all, the performance of a (ML-based) cloud cover parameterization always depends on the accuracy of its inputs, which in turn are affected by other parameterizations in an online setting (e.g., cloud ice/water mixing ratios and specific humidity are modified by ICON's microphysics scheme). Finally, these tasks depend on the correct implementation of the Python-trained NNs into climate model source code (typically written in Fortran). To keep this study tractable, we therefore chose to leave the online implementation for future work; taking the first necessary step of demonstrating a robust offline parameterization, we focus on five questions.



The first key question that we tackle in this study is whether we can train an NN based cloud cover parameterization that is able to emulate high-resolution cloudiness. We then ask the following subquestions: For the sake of generalizability and computational efficiency should we keep the parameterization as local as possible? Or shall we consider non-local effects for improved accuracy? Can we apply this parameterization universally or is it tied to the regions and climatic conditions over which it was trained upon? And can we extract useful physical information from the NN after it has been trained, gaining insight into the interaction between the large-scale (thermo)dynamic state and convective-scale cloudiness?

We first introduce the training data (Sec. \ref{chap:data}) and the NNs (Sec. \ref{chap:nns}), before evaluating regionally (Sec. \ref{chap:regional}) and globally (Sec. \ref{chap:global}) trained networks in their training regime, studying their generalization capability (Sec. \ref{chap:generalizing}) and interpreting their predictions (Sec. \ref{chap:environment}, \ref{Chap:model_errors}).

\section{Data}
\label{chap:sims}
\subsection{ICON High-Resolution Simulations}
\label{chap:data}

The training data consists of coarse-grained data from two distinct ICON storm-resolving model (SRM) simulations. Both simulations provide hourly model output.

The first simulation is a limited-area ICON simulation over the tropical Atlantic and parts of South America and Africa ($10 \degree$S-$20\degree$N, $68 \degree$W-$15 \degree$E). The simulation ran for a bit over two months (December 2013 and August 2016) in conjunction with the NARVAL (NARVALI and NARVALII) expeditions \cite{klocke2017, stevens2019}. The model was initialized at 0 UTC every day and ran for 36 hours. We use the output from the model runs with a native resolution of $\approx$\,2.5\,km. NARVAL data also exists with a higher resolution of $\approx$\,1.2\,km, but it covers a significantly smaller domain (in $4 \degree$S-$18\degree$N, $64 \degree$W-$42 \degree$W). The native vertical grid extends up to 30\,km on 75 vertical layers. \\
The second simulation is a global ICON simulation that ran as part of the QUBICC project. Currently there is a set of hindcast simulations available of which we chose three to work with (hc2, hc3, hc4). Each simulation covers one month (November 2004, April 2005 and November 2005). While the horizontal resolution ($\approx$\,5\,km) is lower than in NARVAL, the vertical grid extends higher (up to 83\,km) on a finer grid (191 layers). 

The two simulations used different collections of parameterization schemes. While the NARVAL simulations were set up to run with ICON's NWP physics package \cite{prill2019}, the QUBICC simulations used the so-called Sapphire physics, developed for SRM simulations and based on ICON's ECHAM physics package as described in \citeA{giorgetta2022}. An overview of the specifically chosen parameterization schemes can be found in Table S1. 
By virtue of their high resolution, both simulations dispensed with parameterizations for convection and orographic/non-orographic gravity wave drag. For microphysics they used the same single-moment scheme, which predicts rain, snow, and graupel in addition to water vapor, liquid water, and ice \cite{doms2011description, seifert2008revised}.
Different schemes were used for the vertical diffusion by turbulent fluxes (NARVAL: \citeA{raschendorfer2001new}, QUBICC: \citeA{mauritsen2007total}), for the radiative transfer (NARVAL: \citeA{barker2003assessing, mlawer1997radiative}, QUBICC: \citeA{pincus2019}), and the land component (NARVAL: \citeA{schrodin2001multi, schulz2015evaluation}, QUBICC: \citeA{raddatz2007will}). The simulations also differed in their cloud cover schemes. The QUBICC simulation assumed to resolve cloud-scale motions, diagnosing a fully cloudy grid cell whenever the cloud condensate ratio exceeds a small threshold and a cloud-free grid cell otherwise. The cloud cover scheme used in NARVAL alternatively produces fractional cloud cover with a diagnostic statistical scheme that combines information from convection, turbulence, and microphysics.

In ICON terminology, the NARVAL simulations ran on an R2B10 and the QUBICC simulations on an R2B9 (horizontal) grid.
Generally speaking, an RnBk grid is a refined spherical icosahedron. 
The refinement is performed by \textbf{i)} dividing its triangle edges into $n$ parts, creating new triangles by connecting the new edge points and by \textbf{ii)} completing $k$ subsequent edge bisections while once more connecting the new edge points after each bisection \cite{giorgetta2018}. 
In between these refinement steps, the position of each vertex is slightly modified using a method called spring dynamics, which improves the numerical stability of differential operators \cite{tomita2001, zaengl2015}.

A key limitation of the data lies in a temporal mismatch between some model output variables from one common time step. This is caused by the sequential processing of some parameterization schemes in the ICON model \cite{giorgetta2018}. For instance, the cloud cover scheme diagnoses cloud cover before the microphysics scheme alters the cloud condensate mixing ratio, which has led to $\approx$\,7\% of the cloudy grid cells in our data to be condensate-free. However, this mismatch should not exceed the fast physics time step in the model, which was set to 40 seconds in the QUBICC and to 24 seconds in the NARVAL simulations. Another limitation of our QUBICC data is that the mixing length in the vertical diffusion scheme was mistakenly set to 1000m instead of 150m, causing unrealistically strong vertical diffusion in some situations.

\subsection{Coarse-Graining}
We now use both NARVAL and QUBICC data to derive training data for our machine learning based cloud cover parameterization.

\begin{figure}
\centering
\hspace*{-1.6cm}\includegraphics[width=1.2\textwidth]{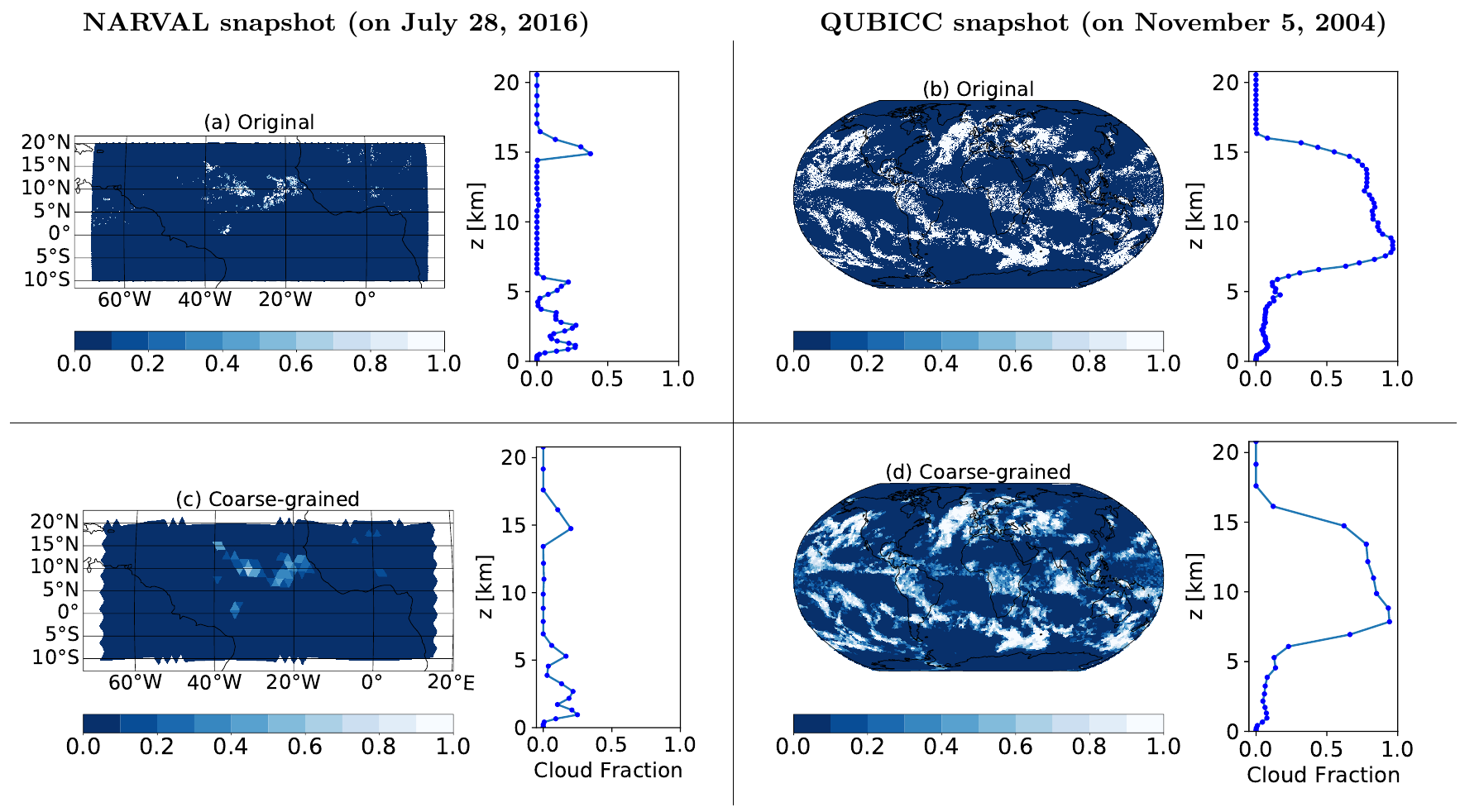}
\caption{Illustration of coarse-graining using the example of cloud fraction. 
Here we show distinct snapshots of the horizontal fields (on a single layer) and vertical profiles (from a single column) from the high-resolution NARVAL and QUBICC simulations (top row) and the corresponding coarse-grained horizontal fields and vertical profiles (bottom row). We coarse-grain the NARVAL/QUBICC data sets horizontally from 2.5\,km/5\,km to 160\,km/80\,km and vertically from 66/87 to 27 layers up to a height of 21\,km. Final coarse-grained grid boxes constitute the training data for the machine learning models.}
\label{fig:cg_figure}
\end{figure}

This requires coarse-graining the data horizontally and vertically to the low-resolution ICON-A grid since we cannot a priori assume that the same (cloud cover) parameterization will work across a very wide range of spatial resolutions. Our goal is to mimic typical inputs of our cloud cover parameterization, which are the large-scale state variables of ICON-A. We design our coarse-graining methodology to best estimate grid-scale mean values, which we use as proxies for the large-scale state variables. Figure \ref{fig:cg_figure} shows an example of horizontal and vertical coarse-graining of cloud cover snapshots from the QUBICC and the NARVAL data set.

We coarse-grain the simulation variables from R2B9 and R2B10 grids to the default R2B4 grid of \citeA{giorgetta2018} with a resolution of $\approx$\,160\,km. To demonstrate the robustness of our machine learning algorithms across typical ICON-A resolutions, we additionally coarse-grain to the low-resolution R2B5 grid used in \citeA{hohenegger2020} with a resolution of $\approx$\,80\,km. Afterwards, we vertically coarse-grain the data to 27 terrain-following sigma height layers, up to a height of 21\,km because no clouds were found above that height. The technical aspects of our coarse-graining methodology can be found in \ref{app:cg}. We now turn towards the specifics of the NNs.

\section{Neural Networks}
\label{chap:nns}
\subsection{Setup}
We set up three general types of NNs of increasing representation power. Each NN follows its own assumption as to how (vertically) local the problem of diagnosing cloud cover is. Choosing three different NN architectures allows us to design a vertically local (cell-based), a non-local (column-based), and an intermediate (neighborhood-based) model type. 
 
The \textbf{(grid-)cell-based model} only takes data from the same grid cell level and potentially some surface variables into account. In that sense, the traditional cloud cover parameterization in ICON-A, being a function of local relative humidity, pressure, and surface pressure, is similarly a cell-based parameterization (with the exception of including the lapse rate in certain situations). Such a local model is very versatile and can be implemented in models with varying vertical grids. \\
The \textbf{neighborhood-based model} has variables as its input that come from the same grid cell and from the ones above and below, and also includes some surface variables. The atmospheric and dynamical conditions in the close spatial neighborhood of the grid cell most likely have a significant influence on cloudiness as well. A grid column undergoing deep convection for instance is very likely to have different cloud characteristics than a grid cell in a frontal stratus cloud \cite{tompkins2005parametrization}. Furthermore, strong subsidence inversions that lead to thin stratocumuli cannot be detected by looking at the same grid cell only. As an example, this dependence of cloudiness on the surroundings has been actualized in \citeA{tompkins2002}. In their study, the sub-grid distribution of total water is described as a function of horizontal and vertical turbulent fluctuations, effects of convective detrainment and microphysical processes. \\
The \textbf{column-based model} operates on the entire grid column at once, and therefore has as many output nodes as there are vertical layers. In a column-based approach we do not have to make any a priori assumptions as to how many grid cells from above and below a given grid cell should be taken into account. Furthermore, surface variables are naturally included in the set of predictors. Coefficients of a multiple linear model fitted to the data suggest that the parameterization of cloud cover is a non-local problem, further motivating the use of a column-based model (see Figure S1). The input-output architecture of these three NN types is illustrated in Figure S2.

We specify three NNs to be trained on the (coarse-grained) NARVAL R2B4 data and three networks to be trained with (coarse-grained) QUBICC R2B5 data. Using data that is coarse-grained to different resolutions allows us to demonstrate the applicability of the approach across resolutions. The primary goal of the NNs trained on NARVAL R2B4 data is to show the ability to reproduce SRM cloud cover from coarse-grained variables, whereas for the globally-trained QUBICC R2B5 NNs it is a versatile applicability and more grid-independence. In this context, the largest differences between the R2B4- and R2B5 models exist in the specification of the neighborhood-based models: \\ 
The set of predictors for the neighborhood-based R2B5 model contains data from the current grid cell and its immediate neighbors (above and below it). On the layer closest to the surface this requires padding to create data from `below'. The vertical thickness of grid cells decreases with decreasing altitude. Therefore, we assume a layer separation of 0 for this artificial layer below, allowing us to fill it with values from the layer closest to the surface. \\ 
The neighborhood-based R2B4 model considers two grid cells above and two below. We did not extend the padding to create another artificial layer, but trained a unique network per vertical layer. This allows for maximum flexibility, discarding input features that are non-existent or constant on a layer-wise basis. Additionally, the R2B4 model has cloud cover from the previous model output time step (1 hour) in its set of predictors.

An overview of the NNs and their input parameters can be found in Table \ref{tab:overview_nns}. The input parameters were mostly motivated by the existing cloud cover parameterizations in ICON-A and the Tompkins Scheme \cite{tompkins2002}. All NNs have a common core set of input features. Choosing varying additional features allows us to study their influence. However, we found that none of these additional features have a crucial impact on a model's performance.
We generally chose as few input parameters as possible to avoid extrapolation situations outside of the training set as much as possible. By doing so, we hope to maximize the generalization capability of the NNs.

\begin{table}
\centering
\caption{Overview of the NNs and their input features. Models N1-N3 are trained on NARVAL R2B4 and models Q1-Q3 on QUBICC R2B5 data. 2D variables (fraction of land/lake, Coriolis parameter and surface temperature) are shaded in purple. More information on the choices and meaning of the features can be found in the SI.}
\rowcolors{2}{}{gray!10}
\label{tab:overview_nns}
\hspace*{-3em}\begin{tabular}{ c l c c c c c c c c c c c c c c }
 & \cellcolor{green!10}{\textbf{NN Type}} & \cellcolor{blue!10}{land} & \cellcolor{blue!10}{lake} & \cellcolor{blue!10}{Cor.} & \cellcolor{blue!10}{$T_s$} & $z_g$ & $q_v$ & $q_c$ & $q_i$ & $T$ & $p$ & $\rho$ & $u$ & $v$ & $clc_{t-1}$ \\
 \cmidrule(lr){2-2}
\cmidrule(lr){3-16}
N1& Cell-based & \CHECK &  & & & \CHECK & \CHECK &  & \CHECK & \CHECK & \CHECK &  &  &  & \\
N2& Column-based &  & \CHECK & & & \CHECK & \CHECK & \CHECK & \CHECK & \CHECK & \CHECK & \CHECK &  &  &\\
N3& Neighborhood-based & & \CHECK & & & \CHECK & \CHECK & \CHECK & \CHECK & \CHECK & \CHECK & \CHECK & & & \CHECK \\
\addlinespace
Q1& Cell-based & \CHECK & & \CHECK & & \CHECK & \CHECK & \CHECK & \CHECK & \CHECK & \CHECK &  & \CHECK & \CHECK &\\
Q2& Column-based & \CHECK & & & & \CHECK & \CHECK & \CHECK & \CHECK & \CHECK & \CHECK &  & & &\\
Q3& Neighborhood-based & & & \CHECK & \CHECK & \CHECK & \CHECK & \CHECK & \CHECK & \CHECK & \CHECK &  & \CHECK & \CHECK &\\
\end{tabular}
\end{table}

\subsection{Training}
In this section we explain the training methodology and the corresponding tuning of the models' and the optimizer's hyperparameters (e.g., model depth, activation functions, initial learning rate). These hyperparameters have a large impact on the potential quality of the NN. The importance of hyperparameter tuning for NN parameterizations was pointed out in  \citeA{ott2020}, and \citeA{yuval2021} proposed its particular need in a real-geography setting.

The choice of hyperparameters for an NN depends on the amount and nature of the training data which in turn depends strongly on the setup. A column-based model in an R2B4 setup trained on NARVAL data can be trained with no more than $1.7 \cdot 10^6$ data samples, using all available data. In contrast, a cell-based model in an R2B5 setup trained on QUBICC data can learn from maximally $4.6 \cdot 10^9$ data samples. Table S2 shows the amount of available training data for every setup. Mainly the coarse-grained QUBICC data had to be (further) preprocessed to a) reduce the size of the data set, b) scale the cloud cover target to a common range, c) normalize the training data, and d) combat the class imbalance of having a relatively large number of cloud-free grid cells in the training data. Steps c) and d) were also necessary for the coarse-grained NARVAL data. The more balanced ratio between cloudy and cloud-free grid cells (which encourages the neural networks to correctly recognize cloudy cells) for d) was achieved by randomly sub-sampling from the cloud-free grid cells. More details on the preprocessing can be found in the SI.


To train the NARVAL R2B4 networks we follow conventional machine learning practices and split the (coarse-grained and preprocessed) R2B4 data into randomly sampled disjoint training, validation and test sets ($78\%/8\%/20\%$ of the data). By randomly splitting the data, we ensure (with a high probability) that the model will see every weather event present in the training data, with the caveat that strongly correlated samples could be distributed across the three subsets. In contrast, for the QUBICC R2B5 models, we focus on universal applicability. We therefore use a temporally coherent three-fold cross-validation split (illustrated in Figure S3). Every fold covers roughly 15 days to make generalization to the validation folds more challenging. We choose 15 days to stay above weather-timescales (so that for instance the same frontal system does not appear in the training and validation folds) and to mitigate temporal auto-correlation between training and validation samples. The validation folds of each split are equally difficult to generalize to, since a part of every month is always included in the training folds. The three-fold split itself lowers the risk of coincidentally working with one validation set that is very conducive to the NN.

After tuning the hyperparameters using the Bayesian optimization algorithm within the SHERPA package \cite{hertel2020sherpa} we found that a common architecture was optimal for the models N1-N3 and Q2. We list the space of hyperparameters we explored in the SI. 
For models Q1 and Q3 we had more training data. To counteract the increase in training time, we increased the batch size to keep a similar amount of iterations per training epoch. After renewed hyperparameter tuning we found a different architecture for models Q1 and Q3. The final choice of hyperparameters for the NNs is shown in Table \ref{tab:neural_network_architectures}. The relatively small size of the NNs (which is comparable to those of \citeA{brenowitz2019}) helps against overfitting the training data and allows for faster training of the networks. By performing systematic optimization of hyperparameters we also found that these networks are already able to capture the functional complexity of the problem. 

\begin{table}
\centering
\caption{Hyperparameters of the NNs and the optimizer}
\rowcolors{2}{}{gray!10}
\label{tab:neural_network_architectures}
\hspace*{-2em}\begin{tabular}{ l  c  c }
 & Models N1-N3 and Q2 & Models Q1 and Q3 \\ 
   \hline
 Hidden layers & $2$ & $3$ \\
 Units per hidden layer & $256$ & $64$ \\
Activation fct. for each layer & ReLU $\rightarrow$ ReLU $\rightarrow$ linear & \begin{tabular}{@{}c@{}}tanh $\rightarrow$ leaky ReLU $(\alpha = 0.2)$ \\ $\rightarrow$ tanh $\rightarrow$ linear\end{tabular} \\
L1, L2 reg. coef. for each layer & None & L1: $4.7 \cdot 10^{-3}$, L2: $8.7 \cdot 10^{-3}$ \\
Batch Normalization & None & After the second hidden layer \\ \addlinespace
Optimizer & N1-N3: Nadam, Q2: Adam & Q1: Adam, Q3: Adadelta \\
$\hookrightarrow$ Initial learning rate & $10^{-3}$ & $4.3 \cdot 10^{-4}$ \\
$\hookrightarrow$ Batch size & N1-N3: $32$, Q2: $128$ & $1028$ \\
$\hookrightarrow$ Maximal number of epochs & N1-N3: $70$, Q2: $40$ & Q1: $30$, Q3: $50$ \\
\end{tabular}
\end{table}

\section{Results}
\subsection{Regional Setting (NARVAL)} \label{chap:regional}
In this section we show the results of the NNs trained and evaluated on the coarse-grained and preprocessed NARVAL R2B4 data (see SI for more details on the preprocessing). For these regionally-trained NNs we define cloud cover as a cloud volume fraction.

\begin{figure}[!htb]
\centering
\hspace*{-0.7cm}\includegraphics[width=1.1\textwidth]{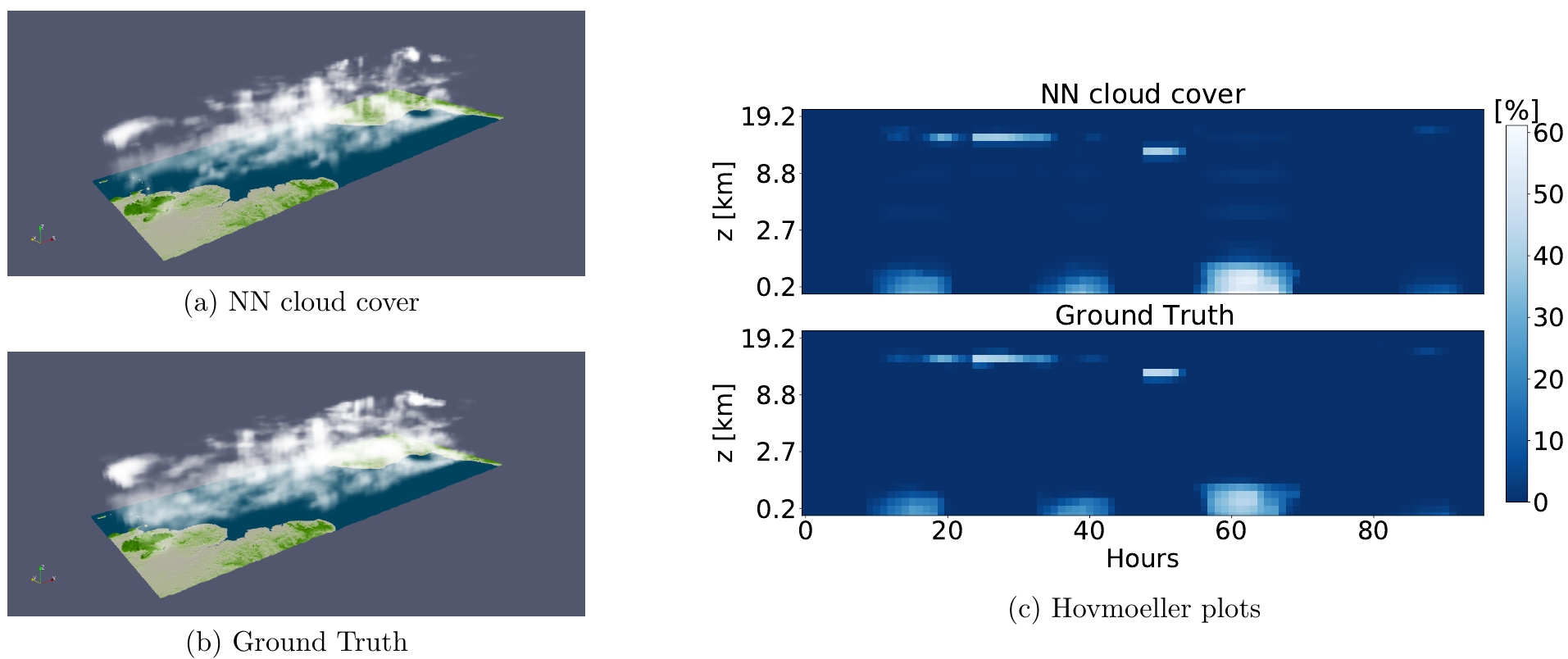}
\caption{The column-based NN trained and evaluated on the coarse-grained NARVAL R2B4 data. Panels a) and b) show cloud cover snapshots with a) displaying the cloud scene as it is estimated by the NN and b) the reference cloud scene from the coarse-grained NARVAL data. Note that some columns over land could not be vertically interpolated due to overlapping topography and are therefore missing in a). The upper plot of panel c) shows the cloud cover predictions of August 1 - August 4, 2016 by the NN in some arbitrary location within the NARVAL region. The plot below depicts the data's actual (coarse-grained) cloud cover. The vertical axis shows average heights of selected vertical layers.}
\label{fig:snapshots_and_hovmoeller}
\end{figure}

The snapshots and Hovmoeller plots of Figure \ref{fig:snapshots_and_hovmoeller} provide visual evidence concerning the capability of the (here column-based) NN to reproduce NARVAL cloud scenes. The ground truth consists of the coarse-grained NARVAL cloud cover fields, which the NN reconstructs while only having access to the set of coarse-grained input features. In the Hovmoeller plots we trace the temporal evolution of cloudiness throughout four days in a randomly chosen grid column of the NARVAL region. Given the large-scale data from the grid column, the NN is able to deduce the presence of all six distinct lower- and upper-level clouds. \\

\begin{table}[!htb]
\centering
\caption{MSEs (in $(\%)^2$) of NARVAL and baseline models evaluated on the coarse-grained and preprocessed NARVAL data}
\rowcolors{2}{}{blue!5}
\label{tab:narval_nn_mses}
\begin{tabular}{l c c c c}
& & & \textbf{Type} & \\
 \cmidrule{3-5}
 & & Cell-based & Column-based & Neighborhood-based \\
\textbf{Neural} & Training set & 15.16 & 1.64 & 0.84  \\
\textbf{networks} & Validation set & 15.18 & 1.78 & 1.00 \\
 & Test set &  15.19 & 1.78 & 1.01 \\
&\setcounter{rownum}{1}\\
\textbf{Baseline} & Constant output model & 109.63 & 92.23 & 86.48 \\
\textbf{models} & Best linear model & 81.71 & 18.56 & 4.79 \\
 & Random forest & 10.40 & 6.15 & 1.73 \\
 & Sundqvist scheme & 51.14 & & \\
\end{tabular}
\end{table}

The models' mean-squared errors (MSEs) (shown in Table \ref{tab:narval_nn_mses}) represent the absolute average squared mismatch per grid cell in percent between the predicted and the true cloud cover. For a given data set $X = \{X_i\}_{i=1}^N$, where for each of the samples $X_i$ the true cloud cover is given by $Y_i$ and the predicted cloud cover by $\hat{Y}_i$, the MSE is defined by

\begin{linenomath*}
\begin{equation}
MSE = \frac1N \sum_{i=1}^N (Y_i - \hat{Y}_i)^2.
\end{equation}
\end{linenomath*}

As opposed to Figure \ref{fig:snapshots_and_hovmoeller}, the MSEs provide more statistically tangible information. The column-based model (which has the largest number of learnable parameters) and the neighborhood-based model (which consists of a unique NN per vertical layer) have lower MSEs than the cell-based model. More trainable parameters allow for the model to adjust better to the ground truth. We also found that by adding more input features (relative humidity, liquid water content, lapse rate and surface pressure) to the cell-based model, we can further decrease its MSE to $\approx$\,5\,$(\%)^2$. On the flip side, every additional input feature bears the risk of impeding the versatile applicability of the model and reducing its capacity to generalize to unseen conditions. By training multiple models of the same type, we verified these MSEs to be robust (varying by $\pm 0.12\,(\%)^2)$. The MSEs for the neighborhood-based model are averaged over all NNs (i.e., one per vertical layer), while the upper-most two layers are left out due to the rare presence of clouds at these altitudes. \\
Our data is temporally and spatially correlated. As a consequence, our division into random subsets for training, validation, and testing leads to very similar MSEs on the respective subsets. And the error on the training set is only slightly smaller than on the validation and test sets. \\
With MSEs being below $16\,(\%)^2$, Table \ref{tab:narval_nn_mses} shows that the NNs are able to diagnose cloud cover better than our baseline models (with the exception of the cell-based random forest).
 These baseline models are fitted to the same normalized data sets as the respective NNs. As our first baseline we evaluate a constant output model, which outputs the average cloud cover. The constant output model's MSE thus also represents the variance of cloud cover in the data. Small differences in the preprocessing of the data for each model type lead to differences in the MSEs of the zero and constant output model. The (multiple) linear model is trained on the data using the ordinary least squares method. For the random forests, we use the default implementation of the RandomForestRegressor in scikit-learn, adjusting the number and the maximum depth of the trees so that the training duration is similar to the NNs. Further adjustments of these two hyperparameters that would further increase or decrease the training durations either reach computational limits or show no decrease in validation loss. While the cell-based random forest actually achieves a lower MSE than the NN, its $\approx$\,$10^5$ larger size ($400$\,GB) makes it impractical to manage. When forced to have a similar storage requirement using the two hyperparameters mentioned above, its MSE ($26.22$) becomes larger than that of the NN. \\ 
 We implemented the Sundqvist scheme as it is described in \citeA{giorgetta2018}. It is a simplified version of the currently implemented (mainly cell-based) ICON-A cloud cover parameterization, because it does not include an adjustment for cloud cover in regions below subsidence inversions over the ocean (see \citeA{mauritsen2019}). We fitted the Sundqvist scheme to the data by doing a grid search over a space of tuning parameters around the values used in the ICON-A model. The grid search yielded a better set of tuning parameters than those found by implementing the scheme as a layer in TensorFlow and optimizing the tuning parameters using gradient descent. To still allow for a differentiation between grid cells over land and ocean, we found optimal sets of tuning parameters for cells that are mainly over land ($\{r_{sat}, r_{0, top}, r_{0, surf}, n\} = \{1.12, 0.3, 0.92, 0.8 \}$) and for cells that are mainly over the sea ($\{r_{sat}, r_{0, top}, r_{0, surf}, n\} = \{1.07, 0.42, 0.9, 1.1 \}$).

\begin{figure}[!htb]
\centering
 \hspace*{-1.6cm}\includegraphics[width=1.2\textwidth]{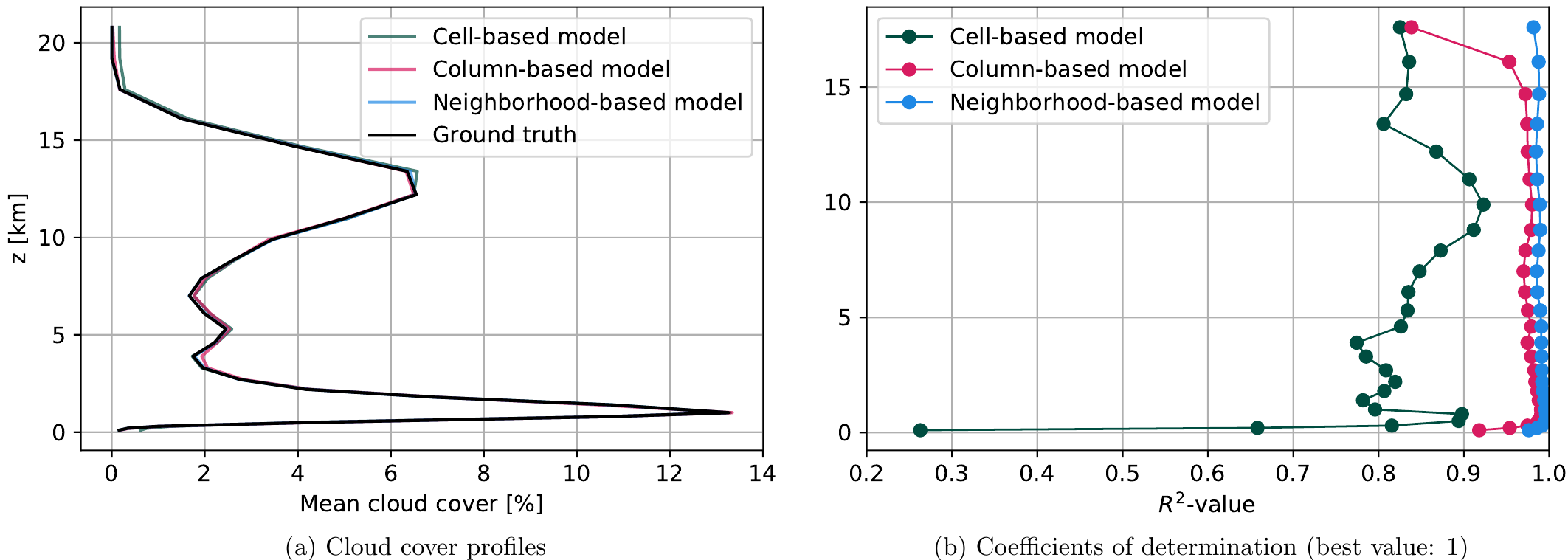}
\caption{Evaluation of the NARVAL R2B4 models on the coarse-grained and preprocessed NARVAL R2B4 data. The three cloud cover maxima of panel a) are located roughly at $1$\,km, $5.3$\,km and $12.2$\,km. The maximal absolute discrepancy between the averaged NN predictions and the ground truth for a given vertical layer is less than $0.5\%$. In panel b), the two upper-most layers are not shown.}
\label{fig:r2b4_combined_mean_clc_and_r2}
\end{figure}

Figure \ref{fig:r2b4_combined_mean_clc_and_r2}a shows that the mean vertical profiles of cloud cover predicted by the NNs closely align with the ``Ground truth'' profile of coarse-grained cloud cover. The profiles feature three maxima that can be attributed to the three modes of tropical convection: shallow, congestus, and deep. Note that in contrast to \citeA{muller2019}, we do find a clear peak for deep convective clouds in the coarse-grained NARVAL and NARVALII data, which could be due to differences in how we define cloudy grid cells (using the cloud cover model output rather than a boolean based on the total cloud condensate mass mixing ratio exceeding $0.1 $g/kg).

In Figure \ref{fig:r2b4_combined_mean_clc_and_r2}b we show the coefficient of determination/$R^2$-value profiles for the different models. For a given vertical layer $l$, the $R^2$-value is defined by

\begin{linenomath*}
\begin{equation}
R^2_l = 1 - \frac{mse_l}{var_l}.
\end{equation}
\end{linenomath*}

For a given vertical layer $l$, $mse_l$
is the mean-squared error between a given model's prediction and the true cloud cover and $var_l$ 
the variance of cloud cover. Clearly, i) $R^2_l \leq 1$, ii) $R^2_l = 1$ implies $mse_l = 0$, and iii) if $R^2_l \leq 0$, then a function always yielding the cloud cover mean on layer $l$ would outperform the model in question.


We see that the neighborhood- and column-based models generally have $R^2$-values exceeding $0.9$, or equivalently $mse_l \leq 0.1 \cdot var_l$. The somewhat lower reproduction skill for the cell-based model concurs with the MSEs found in Table \ref{tab:narval_nn_mses}. 
The models exhibit strongly negative $R^2$-values above $19$\,km and are therefore not shown in the figure, i.e., on these layers a constant-output model would be more accurate than the NNs. The reason for this is that there are almost no clouds above $19$\,km; the variance of cloud cover is not greater than $10^{-4}\,(\%)^2$. Nevertheless, the neighborhood-based model with its unique NN per vertical layer is still able to learn a reasonable mapping at $19.2$\,km, achieving an $R^2$-value of $0.93$. Altogether, we found the mean cloud cover statistics to be independent of how the NNs were initialized prior to training. \\

\subsection{Global Setting (QUBICC)} \label{chap:global}

Having studied the performance of our regionally trained NNs, we now shift the focus to the NNs trained and evaluated on the coarse-grained and preprocessed global QUBICC R2B5 data set. Changing the region as well as the resolution of the training data allows us to conduct studies across these domains in section \ref{chap:environment}.

\begin{table}[!htb]
\centering
\caption{MSEs (in $(\%)^2$) of the NNs trained with a 3-fold cross-validation split on the coarse-grained and preprocessed QUBICC data. We only show the MSEs of the models with the lowest loss on their respective validation folds. Here, the neighborhood-based models comprise one model per split, evaluated on all layers. In parentheses we compute the losses after bounding the model output to the $[0, 100]\%$ interval. The baseline models are trained and evaluated on coarse-grained and preprocessed QUBICC cloud volume fraction data.}
\rowcolors{2}{}{blue!5}
\label{tab:R2B5_mses_qubicc}
\begin{tabular}{l c c c c}
& & & \textbf{Type} & \\
\cmidrule{3-5}
 & & Cell-based & Column-based & Neighborhood-based  \\
 \textbf{Neural} & Cloud volume fraction & 32.77 (28.98) & 8.14 (8.03) & 25.07 (20.46) \\
 \textbf{networks} & Cloud area fraction & 87.98 (80.96) & 20.07 (19.79) & 52.19 (46.61) \\
& & & & \\
  \textbf{Baseline} & Constant output model & 684.51 & 431.28 & 558.28  \\
 \textbf{models} & Best linear model & 401.47 & 97.81 & 297.63 \\
 & Random forest & 25.90 & 161.98 & 54.74 \\
  & Sundqvist scheme & 474.12 & & \\
\hline
   \rowcolor{white}
  \multicolumn{5}{l}{\textit{Due to computational reasons, only $1\%$ of the data (i.e., $\approx$\,$10^7$ samples) was used to}} \\
  \rowcolor{white}
  \multicolumn{5}{l}{\textit{compute the MSE of the Sundqvist scheme.}}
\end{tabular}
\end{table}

\begin{figure}[!htb]
\centering
\includegraphics[width=.9\textwidth]{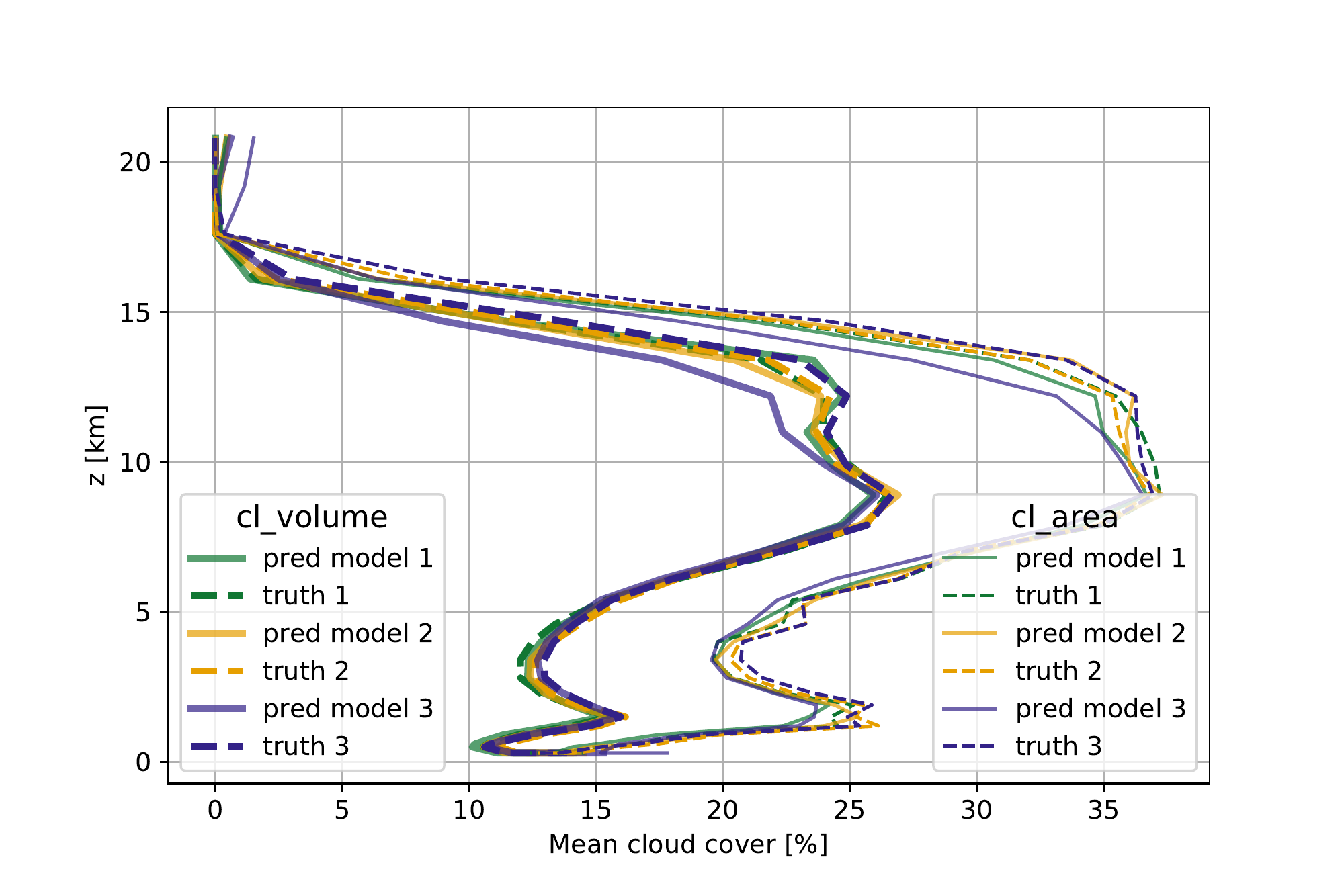}
\caption{The cell-based cloud volume and cloud area fraction models of the 3-fold cross-validation split evaluated on their respective validation sets. The validation losses of the models from split 2 are given in Table \ref{tab:R2B5_mses_qubicc}.}
\label{fig:r2b5_cell-based_validation}
\end{figure}

\begin{figure}[!htb]
\centering
\hspace*{-1.6cm}\includegraphics[width=1.2\textwidth]{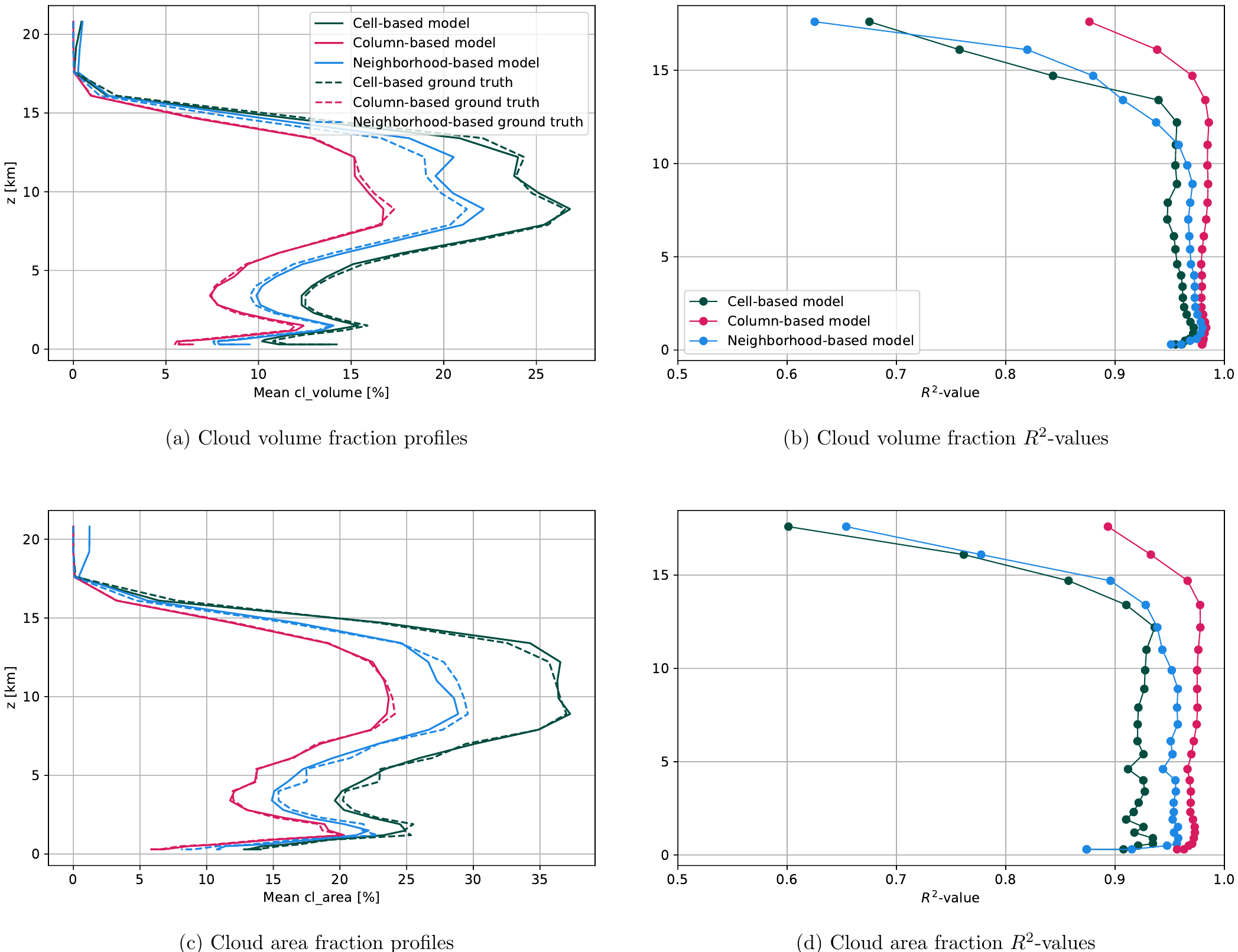}
\caption{Evaluation of QUBICC cloud volume and cloud area models on coarse-grained and preprocessed QUBICC R2B5 data. The layer-wise averaged $R^2$-values of the cell-, column-, and neighborhood-based models shown in b) are (0.94, 0.98, 0.94) and in d) are (0.90, 0.97, 0.93). The ground truth profiles do not match due to differences in preprocessing, especially in how many cloud-free cells were removed from the respective data sets (see SI for more details). The column-based ground truth profile represents the true QUBICC cloud cover profiles since its data was not altered by preprocessing.}
\label{fig:r2b5_qubicc_on_qubicc}
\end{figure}

Table \ref{tab:R2B5_mses_qubicc} shows the performance of the cloud volume and cloud area fraction NNs on their validation folds.
For each model type and each of the three cross-validation splits we trained one NN and then selected the NN that has the lowest MSE on the entire QUBICC data set. Generally, this is also the NN with the lowest loss on its validation set. When comparing Table \ref{tab:R2B5_mses_qubicc} with Table \ref{tab:narval_nn_mses}, we find that QUBICC(-trained) NNs exhibit larger MSEs than NARVAL(-trained) NNs. Causes for the higher MSEs can be attributed to the data now stemming from the entire globe and the higher stochasticity present in the higher resolution R2B5 data. Both of these reasons allow for a larger range of outputs for similar inputs, inevitably increasing the MSE of our deterministic model. Nevertheless, with the exception of the cell-based random forest, we are still well below the MSEs given by our baseline models. However, as in section \ref{chap:regional}, the cell-based random forest requires much more (factor of $\approx$\,$10^6$) memory, and a random forest of similar size to the NN has a larger MSE ($85.86$). The parameters for the Sundqvist scheme were again found using separate grid searches for grid cells that are mainly over land ($\{r_{sat}, r_{0, top}, r_{0, surf}, n\} = \{1.1, 0.2, 0.85, 1.62 \}$) and for grid cells that are mainly over sea ($\{r_{sat}, r_{0, top}, r_{0, surf}, n\} = \{1, 0.34, 0.95, 1.35 \}$). 
In a similar vein, estimating cloud area fraction is a more challenging task than estimating cloud volume fraction. Depending on whether a cloud primarily spans horizontally or vertically, practically any value of cloud area fraction can be attained in a sufficiently humid grid cell. This could explain the increased MSEs of the cloud area fraction models. \\ 
In Table \ref{tab:R2B5_mses_qubicc} we also include bounded losses in parentheses. That means that the NN's cloud cover predictions that are smaller than 0\% are set to 0\% before its MSE is computed. And likewise, predictions greater than 100\% are set to 100\%.
The difference between these two types of losses is relatively small. We can deduce that the NNs usually stay within the desired range of $[0, 100]\%$ without being forced to do so. On average, $76.4\%$ of the predictions of all our QUBICC-trained neural networks in their respective validation sets lie within the $[0, 100]\%$, and $95\%$ of the predictions lie within the slightly larger $[-1, 100]\%$ range.

In Figure \ref{fig:r2b5_cell-based_validation} we show that the local cell-based model -- the model type with the largest MSE -- is still able to reproduce the mean cloudiness statistics of the validation sets that it did not have access to during training. These validation sets each consist of the union of two blocks of 15 days, which is sufficiently temporally displaced from the training data to be above weather timescales. We can see that the validation set bias of the model corresponding to the third split is larger than that of the first two splits. The model from the second split has the overall best performance on the QUBICC data set and is therefore analyzed further in section \ref{chap:generalizing}.

Despite the challenging setting, Figures \ref{fig:r2b5_qubicc_on_qubicc}a and \ref{fig:r2b5_qubicc_on_qubicc}c show that the models are very well able to reproduce the average profiles of cloud volume and cloud area fraction of the global data set. The same holds true for the ability to capture the variance in time and the horizontal for a given vertical layer, which is conveyed by the $R^2$-values being usually well above $0.8$ for all layers below $15$\,km. As in Figure \ref{fig:r2b4_combined_mean_clc_and_r2}, layers above $19$\,km had to be omitted in the $R^2$-plots. When it comes to reconstructing the QUBICC cloudiness, the column-based model with its large amount of adaptable parameters is able to outperform the other two model types. 

After introducing and successfully evaluating both regionally and globally trained networks on their training regimes, we investigate the extent to which we can apply these NNs. 

\subsection{Generalization Capability}
\label{chap:generalizing}
In this section we demonstrate that our globally-trained QUBICC networks can successfully be used to predict cloud cover on the distinct regional NARVAL data set. Furthermore, we show that, with the input features we chose for our NNs, achieving the converse, i.e., applying regionally-trained networks on the global data set, is out of reach.

We note that, beside the regional extent, the QUBICC data covers a different timeframe and was simulated with a different physics package and on a coarser resolution (5\,km) than the NARVAL data (2.5\,km). As opposed to NARVAL's fractional cloudiness scheme, the QUBICC cloud cover scheme diagnosed only entirely cloudy or non-cloudy cells. These differences make the application of NNs trained on one data set to the other data set non-trivial.

\begin{figure}[!htb]
\centering
\hspace*{-1.6cm}\includegraphics[width=1.2\textwidth]{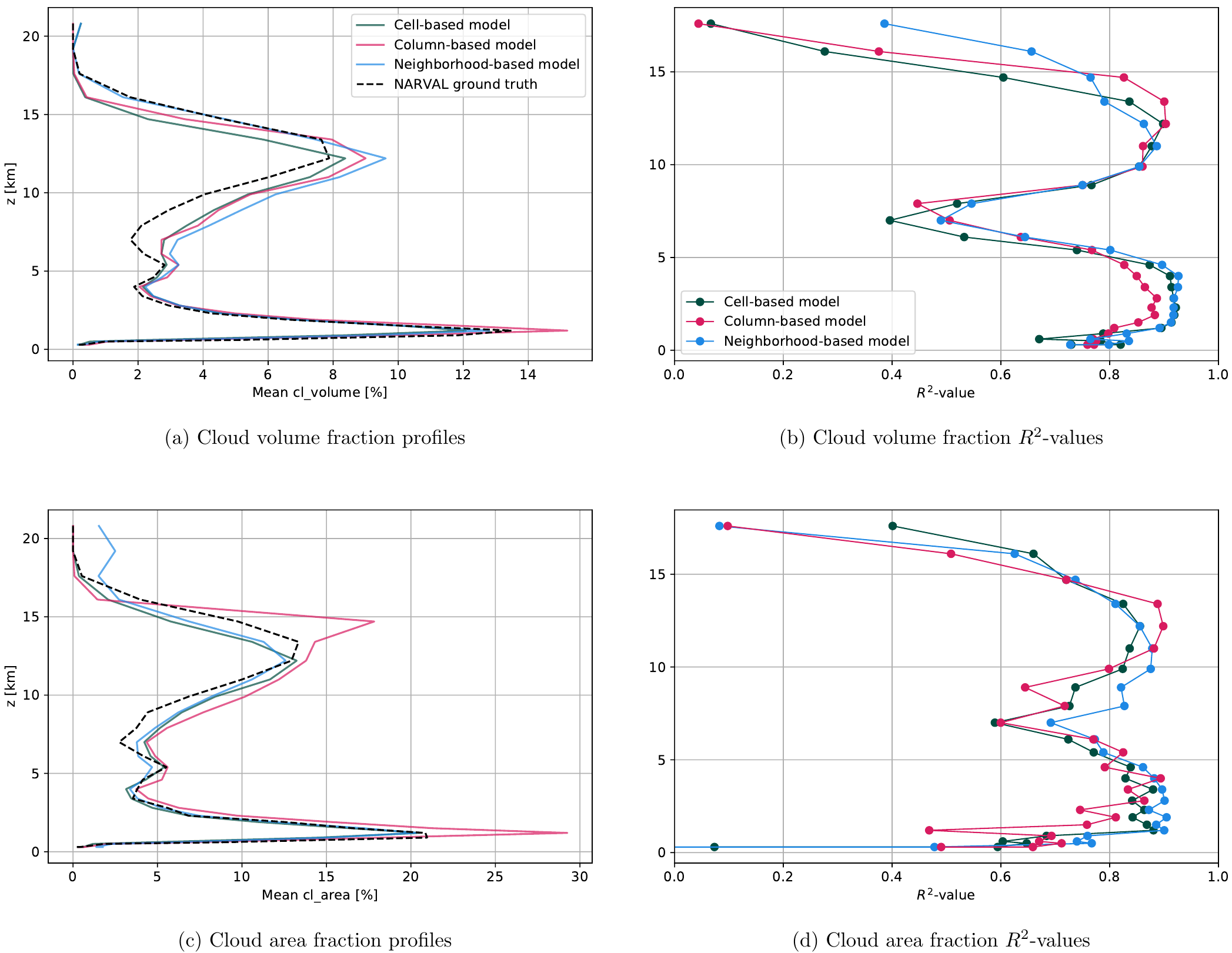}
\caption{Evaluation of QUBICC R2B5 cloud volume and cloud area models on NARVAL R2B5 data. The layer-wise averaged $R^2$-values of the cell-, column-, and neighborhood-based models shown in b) are (0.74, 0.74, 0.79) and in d) are (0.72, 0.71, 0.72).}
\label{fig:r2b5_qubicc_on_narval}
\end{figure}

\textbf{From global to regional} \\
We first study the capability of QUBICC-trained models to generalize to the NARVAL data (see Figure \ref{fig:r2b5_qubicc_on_narval}). We see that the models estimate cloud volume and cloud area fraction quite accurately. This is the case despite the significant differences between QUBICC's and NARVAL's mean vertical profiles of cloud cover. We generally recognize a decrease of $R^2$-value (by $\approx$\,$0.2$) when compared to the models' performance on its training data (Figure \ref{fig:r2b5_qubicc_on_qubicc}). A certain decrease was to be expected with the departure from the training regime. But as the $R^2$-values on average still exceed $0.7$, we find that the models can be applied succesfully to the NARVAL data. In comparison, the Sundqvist scheme we tuned on the QUBICC R2B5 data has a layer-wise averaged $R^2$-value of $-0.54$/$0.29$ for cloud volume/area fraction on the NARVAL data, but only if we discard the surface-closest layer.

However, there is a significant bias affecting all three NN types, namely consistent overprediction of both cloud volume and cloud area fraction between $6$ and $9$\,km. In this altitude range, this is visible in all four plots, either through the mismatch in mean cloud cover or the dip in $R^2$-value. This behavior will be further investigated in section \ref{Chap:model_errors}. Another minor bias is a slightly poorer generalization of the column-based model to the NARVAL data (see e.g., Figure \ref{fig:r2b5_qubicc_on_narval}c). We can understand this as a sign of overfitting if we also take into account that the column-based model showed a higher skill on the training data than the other two model types.

\textbf{From regional to global} \\
We have seen that the NNs are able to reproduce the cloud cover distribution of the storm-resolving NARVAL simulation, limited to its tropical region. 
We coarse-grain the QUBICC data to the same R2B4 grid resolution that the NARVAL NNs were trained with. This helps us to investigate to what extent the NNs can actually generalize to out-of-training regimes. We focus on the tropics first, extending the evaluation from the NARVAL region (68W-15E, 10S-20N) to the entire tropical band (23.4S-23.4N). Note that the QUBICC data shows a much stronger presence of deep convection and a weaker presence of shallow and congestus-type convection. Nevertheless, the NNs are able to reproduce the general structure of the mean cloud cover profile, in particular the peak due to deep convection. The flattened peak of shallow convection is most accurately represented by the neighborhood-based model, while the weakened congestus-type convection is reproduced by both the neighborhood- and the column-based models. 

\begin{figure}[!htb]
\centering
\hspace*{-1.6cm}\includegraphics[width=1.2\textwidth]{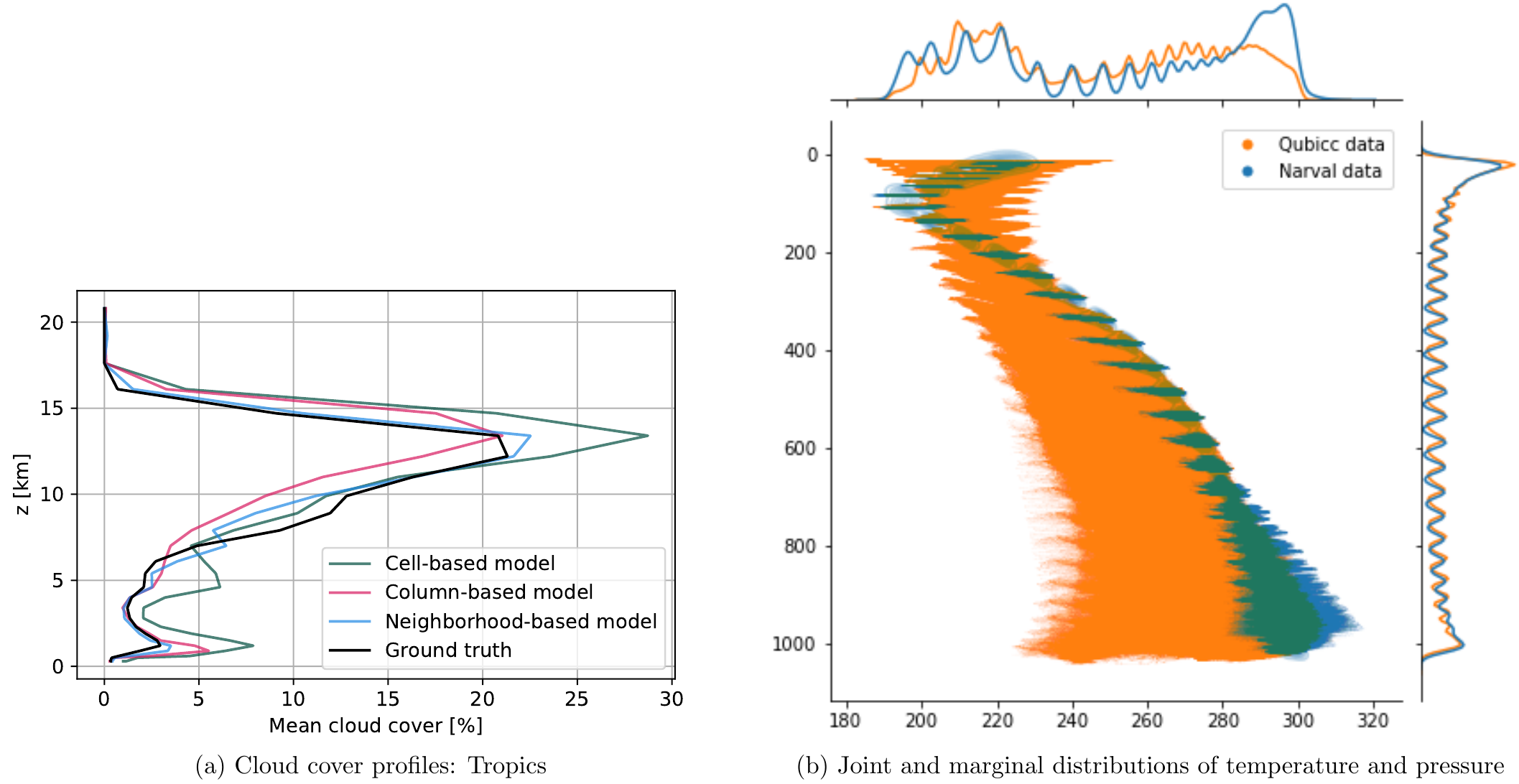}
\caption{Panel a): Evaluation of NARVAL R2B4 models (NARVAL region: 68W-15E, 10S-20N) on QUBICC R2B4 data over the tropical zone (23.4S - 23.4N). We plot the means over 10 days (Nov. 20 - Nov. 29, 2004). Different NNs of the same type produce consistent mean vertical cloudiness profiles ($\pm 1 \%$). The layer-wise averaged $R^2$-values below $15$\,km of the cell-, column-, and neighborhood-based models are (-0.88, 0.29, 0.67), and within the upper troposphere (between $6$ and $12$\,km) they are (0.72, 0.62, 0.84). Panel b): Joint distribution of temperature and pressure in NARVAL R2B4 and QUBICC data. On the margins we see the univariate distributions of temperature and pressure. The jagged structure emerges from the underlying coarse vertical grid.}
\label{fig:r2b4_narval_on_global}
\end{figure}

However, the NNs are not able to generalize to the entire globe. To show this, we use two column-based models as an example. Looking at Figure S4, we can see that they are unable to reproduce mean cloudiness statistics over the region covering the Southern Ocean and Antarctica. In addition, models with the same architecture produce entirely different cloudiness profiles. In this polar region, the NNs are evidently forced to extrapolate to out-of-training regimes and are thus unable to produce correct or consistent predictions. Let us look exclusively at the univariate distributions of the QUBICC input features (those for temperature and pressure are plotted on the margins of Figure \ref{fig:r2b4_narval_on_global}b). Then we can see that their values are usually covered by the distribution of the NARVAL training data. Only their joint distribution reveals that a large number of QUBICC samples exhibit combinations of pressure and temperature that were not present in the training data. For instance, temperatures as cold as 240K never occur in tandem with pressure values as high as 1000\,hPa in the tropical training regime of the NARVAL data. This circumstance is particularly challenging for the neighborhood- and column-based models. This is because the input nodes in these two NARVAL model types correspond to specific vertical layers. So the NNs have to extrapolate when facing (during training) unseen input feature values on any vertical layer, such as in our example cold temperatures on a vertical layer located at around 1000\,hPa.

 
In this section, we demonstrated that the QUBICC NNs can be used on NARVAL data, while in our setup the converse is not feasible. This begs the question: In which way do these NNs differ and have they actually learned a meaningful dependence of cloud cover on the thermodynamic environment? 

\subsection{Understanding the Relationship of Predicted Cloud Cover to Its Thermodynamic Environment}
\label{chap:environment}
In this section, our goal is to dig into the NNs and understand which input features drive the cloud cover predictions. We furthermore want to uncover similarities and differences between the NARVAL- and QUBICC-trained NNs that help understand differences in their generalization capability. 

NNs are not inherently interpretable, i.e., we cannot readily infer how the input features impacted a given prediction by simply looking at the networks' weights and biases. Instead, we need to use an \textit{attribution method} that uses an explanation method built on top of the NN \cite{ancona2019}.
Within the class of attribution methods, few are adapted for regression problems. A common choice (see e.g., \citeA{brenowitz2020}) is to use gradient-based attribution methods. However, these methods may not fairly account for all inputs when explaining a model's prediction \cite{ancona2019}. Additionally, gradient-based approaches can be strongly affected by noisy gradients \cite{ancona2019} and generally fail when a model is `saturated', i.e., when changes in the input do not lead to changes in the output \cite{shrikumar2017}. \\

Instead we approximate Shapley values for every prediction using the SHAP (SHapley Additive exPlanations) package \cite{lundberg2017}. The computation of Shapley values is solidly founded in game theory and the Shapley values alone satisfy three `desirable' properties \cite{lundberg2017}. Shapley values quantify the influence of how an input feature moves a specific model prediction away from its \textit{base value}, defined as the expected output. The base value is usually an approximation of the average model output on the training data set. With Shapley values, the difference of the predicted output and the base value is fairly distributed among the input features \cite{molnar2020}. A convenient property is that one can recover this difference by summing over the Shapley values (`efficiency property').  \\
The DeepExplainer within the SHAP package is able to efficiently compute approximations of Shapley values for deep NNs \cite{lundberg2017}. SHAP also comes with various visualization methods, which allow us to aggregate local sample-based interpretations to form global model interpretations.


We now show how we use SHAP to compare the way NARVAL (R2B4)- and QUBICC (R2B5)-trained networks arrive at good predictions. We focus on the column-based (cloud volume fraction) models. These are uniquely able to uncover important non-local effects, have the largest number of input features to take into account and have on average the lowest MSEs in their training regimes (taking into account both Table \ref{tab:narval_nn_mses} \textit{and} \ref{tab:R2B5_mses_qubicc}).

We collect local explanations on a sufficiently large subset of the NARVAL R2B5 data. For this, we compute the base values by taking the average model predictions on subsets of the respective training data sets. A necessary condition for the base value is that it approximates the expected NN output (on the entire training set) well. We found that $\approx 10^4$ QUBICC samples are sufficient for the average NN prediction to converge. Therefore, we used this size for the random subsets of the QUBICC and of the smaller NARVAL training set as well. We showed that on the NARVAL R2B5 data set, the QUBICC models are able to reconstruct the mean vertical profile with high $R^2$-values (Figure \ref{fig:r2b5_qubicc_on_narval}).
Impressively, the column-based version of our NARVAL R2B4 models also makes successful predictions on the NARVAL R2B5 data set (with an average $R^2$-value of $0.93$; Figure S5) despite the doubling of the horizontal resolution.

The size of the subset of NARVAL R2B5 data ($\approx$\,$10^4$ samples) is chosen to be sufficiently large to yield robust estimates of average absolute Shapley values. Averaging the absolute Shapley values over many input samples measures the general importance of each input feature on the output. An input feature with a large average absolute Shapley value contributes strongly to a change in the model output. It on average increases or decreases the model output by precisely this value.

\begin{figure}[!htb]
\centering
\hspace*{-3.2cm}\includegraphics[width=1.4\textwidth]{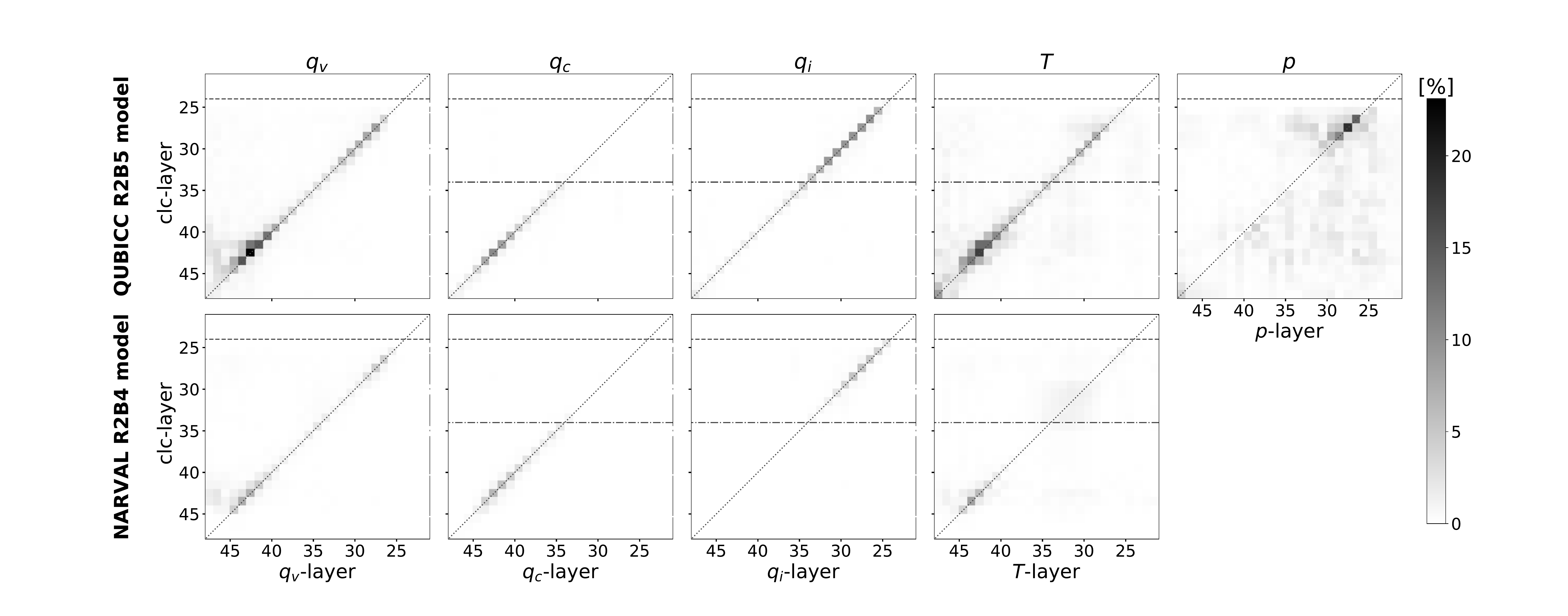}
\caption{Average absolute SHAP values of the QUBICC R2B5 and the NARVAL R2B4 column-based models when applied to the same, sufficiently large subset of the NARVAL R2B5 data. We use the conventional ICON-A numbering of vertical layers from layer 21 (at a height of $\approx$\,$20.8$\,km) decreasing in height to layer 47, which coincides with Earth's surface. The dashed line shows the tropopause, here at $\approx$\,15\,km, the dash dotted line shows the freezing level (i.e., where temperatures are on average below $0$ degrees C), here at $\approx$\,5\,km. Tests with four different seeds show that the pixel values are robust (the absolute values never differ by more than $0.55\%$). The input features that are not shown exhibit smaller absolute SHAP values ($\rho < 1.8\%$, $p < 1.5\%$, $z_g < 0.7\%$, \textit{land}/\textit{lake} $< 0.1\%$) everywhere and are thus omitted.}
\label{fig:average_absolute_SHAP_values}
\end{figure}


The absolute SHAP values (Figure \ref{fig:average_absolute_SHAP_values}) suggest that both models learned a remarkably local mapping, with a clear emphasis on the diagonal (especially above the boundary layer). That means that the prediction at a given vertical layer mostly depends on the inputs at the same location. The models have learned to act like our cell- or neighborhood-based models without human intervention. \\
The input features have a larger influence in the QUBICC model than they do in the NARVAL model. We can also see this phenomenon, if we use a similar base value for both models (see Figure S6). This is most likely due to the fact that the QUBICC model was exposed to a wide variety of climatic conditions across the entire globe during training, resulting in a greater variance in cloud cover. The NN is thus used to deviate from the average cloud cover, putting more emphasis on its input features, and consequently causing larger Shapley values. \\
Both models take into account that in the boundary layer the supply of moisture $q_v$ from below in combination with temperature anomalies that could drive convective lifting influence the sub-grid distribution of cloud condensates and henceforth cloud cover. Such a non-local mixing due to updrafts presents limitations for purely local parameterizations. In the boundary layer (which we define to be at below 1\,km), temperature $T$ and specific humidity $q_v$ are found to be the most important variables (having the largest sum of absolute SHAP values) for the NNs. 
Higher in the troposphere, the local amount of moisture has a significant impact on cloud cover. Specific cloud liquid water content $q_c$ is a major predictor of cloud cover below the freezing level, while specific cloud ice content $q_i$ is a major predictor of cloud cover above the freezing level. 
In contrast to the global QUBICC model, the tropical NARVAL model only considers the impact of $q_i$ at sufficiently high altitudes, which allow for the formation of cloud ice. The QUBICC model also learned to place more emphasis on $T$ and $q_v$ in the lower troposphere and pressure $p$ in the higher troposphere than the NARVAL model. \\
Generally, the most important variables above the boundary layer and below the freezing level are temperature $T$ (for the QUBICC model) and cloud water $q_c$ (for the NARVAL model). Above the freezing level, the QUBICC model emphasizes pressure $p$ most, while the NARVAL model learns a similar impact of $T$, $q_i$ and $p$ (not shown).  
Due to the Clausius-Clapeyron relation, relative humidity depends most strongly on temperature. Taking into account that throughout the troposphere relative humidity is the best single indicator for cloud cover \cite{walcek1994}, this is a likely explanation for the models' large emphasis on temperature.

After using SHAP to illustrate which features drive the (column-based) NN predictions, we use the same approach to understand the source of a specific generalization error of the QUBICC NNs (Figure \ref{fig:r2b5_qubicc_on_narval}). 

\subsection{Understanding Model Errors}
\label{Chap:model_errors}

In this section, our goal is to understand the source of flawed NN predictions. We want to analyze what type of dependence on which input features is most responsible for erroneous predictions. This type of analysis reveals differences in the (NN-learned) characteristics of the training data set and a data set to which an NN is applied to.

In the evaluation of the QUBICC (R2B5) cloud volume fraction models on NARVAL R2B5 data (Figure \ref{fig:r2b5_qubicc_on_narval}) we have seen a pronounced dip in performance ($R^2 \leq 0.8$ for all models) on a range of altitudes between 6 and 9\,km. The dip was accompanied by an overestimation of cloud cover (relative error $> 15\%$). We specifically focus on explaining the bias at 7\,km. The vertical layer that corresponds to this altitude is the 32\textsuperscript{nd} ICON-A layer. On layer 32, the $R^2$-values are minimal ($R^2 \leq 0.5$ for all models) making it arguably the largest tropospheric generalization error of the models. However, the method we employ here can be used to understand other generalization errors as well.

The NARVAL (R2B4) models are perfectly able to make predictions on NARVAL R2B5 data on layer 32 (Figure S5), making it a suitable benchmark model. As in the previous section we use SHAP on the column-based models. In order to be able to compare Shapley values corresponding to certain features individually, we follow the strategy outlined in \ref{app:shap}.

\begin{figure}[!htb]
\centering
\hspace*{-3.8cm}\includegraphics[width=1.5\textwidth]{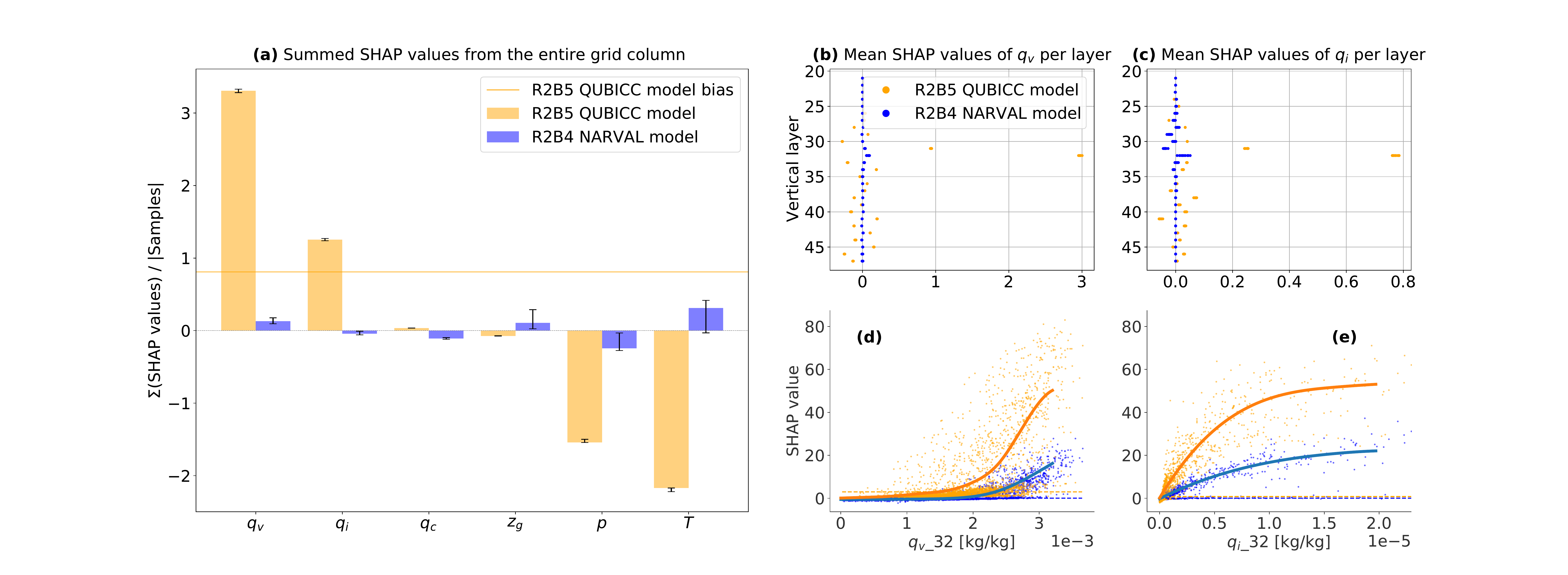}
\caption{SHAP/Shapley value statistics per input feature for cloud cover predictions on vertical layer 32 (at $\approx$\,7\,km) of the column-based models with a focus on $q_v$ and $q_i$ in \textbf{(b)}-\textbf{(e)}. Input features the models have not in common are neglected. As in Figure \ref{fig:average_absolute_SHAP_values}, the Shapley values for both models are computed on the same sets of $10^4$ random NARVAL R2B5 samples (using ten different seeds). 
\textbf{(a)}: The sum of average SHAP values over all vertical layers. The black lines show the range of values (min/max). The absolute QUBICC R2B5 model bias (of $0.95\%$) on layer 32 (cf. Figure \ref{fig:r2b5_qubicc_on_narval}a) can approximately be recovered by summing over all orange values (which yields $0.81\%$). \textbf{(b)}, \textbf{(c)}: The vertical profiles of SHAP values for $q_v$ and $q_i$ for all ten seeds. In the SHAP dependence plots \textbf{(d)}, \textbf{(e)} we zoom in on the features with the largest SHAP values ($q_i$ and $q_v$ of layer 32).
\textbf{(d)}, \textbf{(e)}: Each dot corresponds to one NARVAL R2B5 sample. The lines show smoothed  conditional expectations computed over all seeds. The dashed lines show the average SHAP value of the input features $q_v$ and $q_i$ on layer 32 whose values can also be found in \textbf{(b)} and \textbf{(c)}.}
\label{fig:shap_clc_32_all_plots}
\end{figure}

Figure \ref{fig:shap_clc_32_all_plots}a shows the influence of each input feature from the entire grid column on the average model output on layer 32. We find that the QUBICC model bias is driven by $q_v$ and $q_i$. Compared to the NARVAL model, the QUBICC model clearly overestimates the impact of these two variables. This impact is dampened somewhat by a net decreasing effect of $p$ and $T$ on the cloud cover predictions. 
In the NARVAL model the impact of these features is much less pronounced. The reason is probably once again that the model has not learned the need for deviating much from the base value in its tropical training regime.

When investigating the vertical profile of Shapley values in Figures \ref{fig:shap_clc_32_all_plots}b and c we find that the local values have the largest effect on cloud cover. This local importance is also corroborated by Figure \ref{fig:average_absolute_SHAP_values}. We can zoom in and look at the more precise conditionally-averaged functional dependence
of $clc$\_32 on these local $q_i$\_32 and $q_v$\_32 variables (Figures \ref{fig:shap_clc_32_all_plots}d and e). We find the two functions to be very similar, albeit differing in their slope. The QUBICC model quickly increases cloud cover with increasing values of $q_i$\_32 and $q_v$\_32. The QUBICC model's large emphasis on $q_i$\_32 could be a relict from the cloud cover scheme in the native QUBICC data. This scheme had set cloud cover to $100\%$, whenever the cloud condensate ratio had exceeded a given threshold.


\section{Summary}




In this study we develop the first machine learning based parameterization for cloud cover based on the ICON model and deep NNs. We train the NNs with coarse-grained data from regional and global storm-resolving model simulations with real geography. We demonstrate that in their training regime, the NNs are able to learn the sub-grid scale cloud cover from large-scale variables (Figures \ref{fig:r2b4_combined_mean_clc_and_r2}, \ref{fig:r2b5_qubicc_on_qubicc}). Additionally we show that our globally trained NNs can also be successfully applied to data originating from a regional simulation that differs in many respects (e.g., its physics package, horizontal/vertical resolution, and time frame; Figure \ref{fig:r2b5_qubicc_on_narval}). Using SHAP we compare regionally and globally trained NNs to understand the relationship between predicted cloud cover and its thermodynamic environment and vertical structure (Figure \ref{fig:average_absolute_SHAP_values}). We are able to uncover that specific humidity and cloud ice are the drivers of one NN's largest tropospheric generalization error (Figure \ref{fig:shap_clc_32_all_plots}). \\
We implement three different types of NNs in order to assess the degree of (vertical) locality and the amount of information they need when it comes to the task of diagnosing cloud cover. We find that by enforcing more locality, the performance of the NN suffers on its training set (Figures \ref{fig:r2b4_combined_mean_clc_and_r2}, \ref{fig:r2b5_qubicc_on_qubicc}). However, the more local cell- and neighborhood-based NNs show slightly fewer signs of overfitting the training data (Figure \ref{fig:r2b5_qubicc_on_narval}). Generally we find that none of the three types clearly outperforms the other two types and that the potentially non-local model in actuality also mostly learned to disregard non-local effects (Figure \ref{fig:average_absolute_SHAP_values}).
Overall, the neighborhood-based model trained on the global QUBICC data (Q3) is most likely the preferable model. It has a good accuracy on the training data, the lowest generalization error on the NARVAL data, is low-dimensional, easy to implement and cross-model compatible. The last point refers to the fact that (unlike the column-based model) it is not tied to the vertical grid it was trained on. \\
Furthermore, the NNs are trained to differentiate between cloud volume and cloud area fraction, which are distinct interpretations of cloud cover. We found cloud area fraction to be a somewhat more difficult value to predict. The shape of a cloud, which determines its cloud area fraction, is harder to extract from grid-scale averaged thermodynamic variables.
We agree with \citeA{brooks2005} that a distinction between these two concepts of cloud cover would be expedient inside a general circulation model for two reasons: First, both interpretations are used in the microphysics and radiation schemes. Second, depending on the interpretation, cloud cover can differ significantly (Figure \ref{fig:cloud_cover_vs_area}).

The natural next step will be to implement and evaluate the machine learning based parameterization for cloud cover in the ICON model. In such an ICON-ML model, the machine learning based parameterization would substitute the traditional cloud cover parameterization. The NN predictions for cloud area and cloud volume fraction would be used as parameters for the radiation and microphysics parameterizations, depending on which interpretation is most appropriate in each case. Preliminary online simulations covering one QUBICC month (not shown) demonstrate the potential of our neighborhood-based NN parameterization as it is (a) able to process its input variables from the coarse-scale distributions while (b) pushing the statistics of, e.g., the cloud water mixing ratio, to that of the (coarse-grained) high-resolution statistics as desired. However, as we discussed in the introduction, more work is required to create an ICON-ML model that produces accurate and robust results.

The presence of condensate-free clouds in the training data shows inaccuracies that are present both in the NARVAL and the QUBICC training data. These could have been avoided by introducing targeted multiple calls to the same parameterization scheme in the high-resolution model that generated the data. However, we emphasize that the machine learning approach is general enough that if the data were generated more carefully then our approach would still work. \\
Our regionally-trained networks are not able to generalize to the entire globe. Similar difficulties might arise when applying our globally-trained networks to a very different climate \cite{rasp2018pnas}. In practice, this would require us to filter out data samples which the NN cannot process in a meaningful way. Alternatively, one could train the NNs with climate-invariant features only, eliminating the need of ever extrapolating to out-of-training distributions \cite{beucler2021}. By additionally using causal discovery methods to guide their selection, one would most likely arrive at a more rigorous and physically consistent set of input features \cite{nowack2020, runge2019}. Another useful modification to our NNs would be to add a method that allows us to estimate the uncertainty associated with a prediction, e.g., either by adding dropout \cite{gal2016} or by implementing the NNs as Bayesian NNs.



From a climate science perspective, instead of diagnosing cloud cover from large-scale variables directly, one could also train an NN to output parameters specifying distributions for sub-grid scale temperature and moisture.
Cloud cover could then be derived from these distributions (see \textit{statistical cloud cover schemes} in e.g., \citeA{stensrud2009, tompkins2002}). By reusing the distributions for other parameterizations as well, we could increase the consistency among cloud parameterizations. However, this approach would require us to make assumptions concerning the general form of these distributions \cite{larson2017} and we leave this for future work.

Overall, this study demonstrated the potential of deep learning combined with high-resolution data for developing parameterizations of cloud cover.

\appendix
\section{Coarse-graining methodology} \label{app:cg}
Our goal is to to best estimate grid-scale mean values. Ideally, we would derive the large-scale grid-scale mean $\bar{S}$ of a given variable $S$ by integrating over the grid cell volume $V \subseteq \mathbb{R}^3$. In practice, we compute a weighted sum over the values $S_{i,j}$ of all high-resolution grid cells $H$. Here, $i$ is the horizontal and $j$ is the vertical index of a high-resolution grid cell. We define the weights $\alpha_{i,j} \in [0, 1]$ as the fraction of $V$ that a high-resolution grid cell indexed by $(i,j)$ fills. This is a basic discretization of the integral.

To make this term easier to compute in practice, we introduce another approximation. Instead of computing $\alpha_{i,j}$ directly, we split it into the fraction of the horizontal area of $V$ (denoted by $\gamma_i \in [0, 1]$) \textit{times} the fraction of the vertical thickness of $V$ (denoted by $\beta_j \in [0, 1]$) that the high-resolution grid cell indexed by $(i,j)$ fills. We first compute the weights $\gamma_i$ and the weighted sum over the horizontal indices $i$ (horizontal coarse-graining). Only afterwards do we compute the weights $\beta_j$ and the weighted sum over the vertical indices $j$ (vertical coarse-graining).
 
Note that this is indeed an approximation. The geometric heights and vertical thicknesses of grid cells in $H$ on a specific vertical layer $j$ do not need to match exactly. These slight differences are lost when horizontally coarse-graining to fewer grid boxes. Therefore, the second approximation is an approximation because we $\textbf{i)}$ compute the vertical overlap $\beta_j$ \textit{after} we horizontally coarse-grain the grid cells and $\textbf{ii)}$ work on a terrain-following height grid which allows for vertical layers of varying heights over mountaineous land areas. Over ocean areas, where the height levels have no horizontal gradient, this simplification in the computation of the weights has no disadvantage. 


In short, let $\alpha_{i,j}, \beta_j, \gamma_i \in [0, 1]$ be the weights describing the amount of overlap in volume/vertical/horizontal between the high-resolution grid cells and the low-resolution grid cell. We then calculate the large-scale grid-scale mean as the weighted sum of high-resolution variables

\begin{linenomath*}
\begin{equation}
\bar{S} \equiv \frac1{\vert V \vert} \int_{V} S dx \approx \sum_{(i,j)\in H} \alpha_{i,j} S_{i,j} \approx \sum_{(i,j) \in H} \beta_j \gamma_i S_{i,j}.
\end{equation}
\end{linenomath*}

\begin{figure}
\centering
\hspace*{-5.5em}\includegraphics[width=1.3\textwidth]{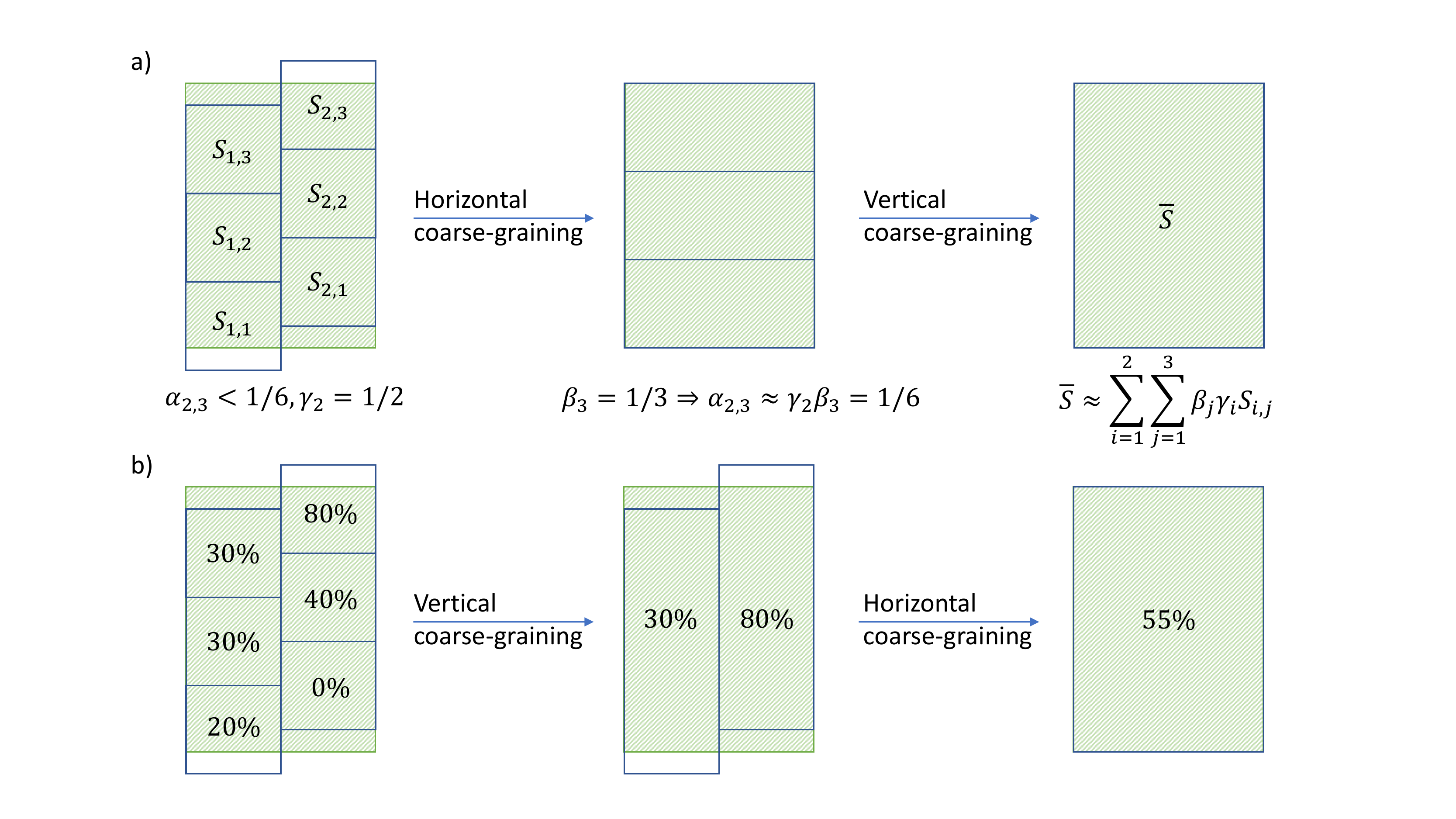}
\caption{Sketch of our general coarse-graining methodology in panel a) and for cloud area fraction in panel b). We picture a vertical slice through two grid columns. For simplicity we assume that the grid boxes all have the same depth. The greenly hatched area depicts a coarse-scale grid box $V$. Panel a): Due to our approximation the weight $\alpha_{2,3}$ for the value in grid box $S_{2,3}$ is $1/6$ and therefore larger than it were  without the sequential horizontal and vertical coarse-graining steps. Panel b): In the vertical range of $V$ we vertically coarse-grain cloud cover values according to a maximum overlap assumption before we coarse-grain in the horizontal.}
\label{fig:visualize_cg}
\end{figure}

We also illustrate our approach in panel a) of Figure \ref{fig:visualize_cg}. \\
The use of spring dynamics in between model grid refinement steps allows for the presence of fractional horizontal overlap $\gamma_i$. As our method for horizontal coarse-graining we choose the first order conservative remapping from the CDO package \cite{schulzweida_uwe_2019_3539275}, which is able to handle fractional overlap and the irregular ICON grid to coarse-grain to and from.

There are locations where the low-resolution grid cells that are closest to Earth's surface extend significantly further downwards than the high-resolution grid cells. This is due to topography that can only be seen at fine scales and makes it difficult to endue these low-resolution grid cells with a meaningful average computed from the high-resolution cells. We therefore omit these grid cells during coarse-graining. This issue is present only in scattered, isolated grid cells over land and it affects a small fraction of all grid cells ($0.2$\%) and columns ($4.7$\%). So it does not pertain entire regions, which would decrease the scope and quality of the data set. While horizontally coarse-graining NARVAL data, we analogously omit low-resolution grid cells that are not located entirely inside the NARVAL region.

To derive the cloud area fraction $C$ we cannot start by coarse-graining horizontally. We first need to utilize the high-resolution information on whether the fractional cloud cover on vertically consecutive layers of a low-resolution grid column overlaps or not. Therefore, we first vertically coarse-grain cloud cover to a grid that would -- after subsequently horizontally coarse-graining -- resemble the ICON-A grid as much as possible. For the first step, we assume maximum overlap as the level separation of vertical layers is relatively small. We thus calculate the coarse-grained cloud area fraction $\overline{C}$ as the sum of the vertically maximal cloud cover values $\max_j\{C_{i,j}\}$ weighted by the horizontal grid cell overlap fractions $\gamma_i$

\begin{linenomath*}
\begin{equation} \label{cloud_area_fraction}
\overline{C} = \sum_{(i,j) \in H} \gamma_i \max_j\{C_{i,j}\}.
\end{equation}
\end{linenomath*}

\begin{figure}
\centering
\includegraphics[width=\textwidth]{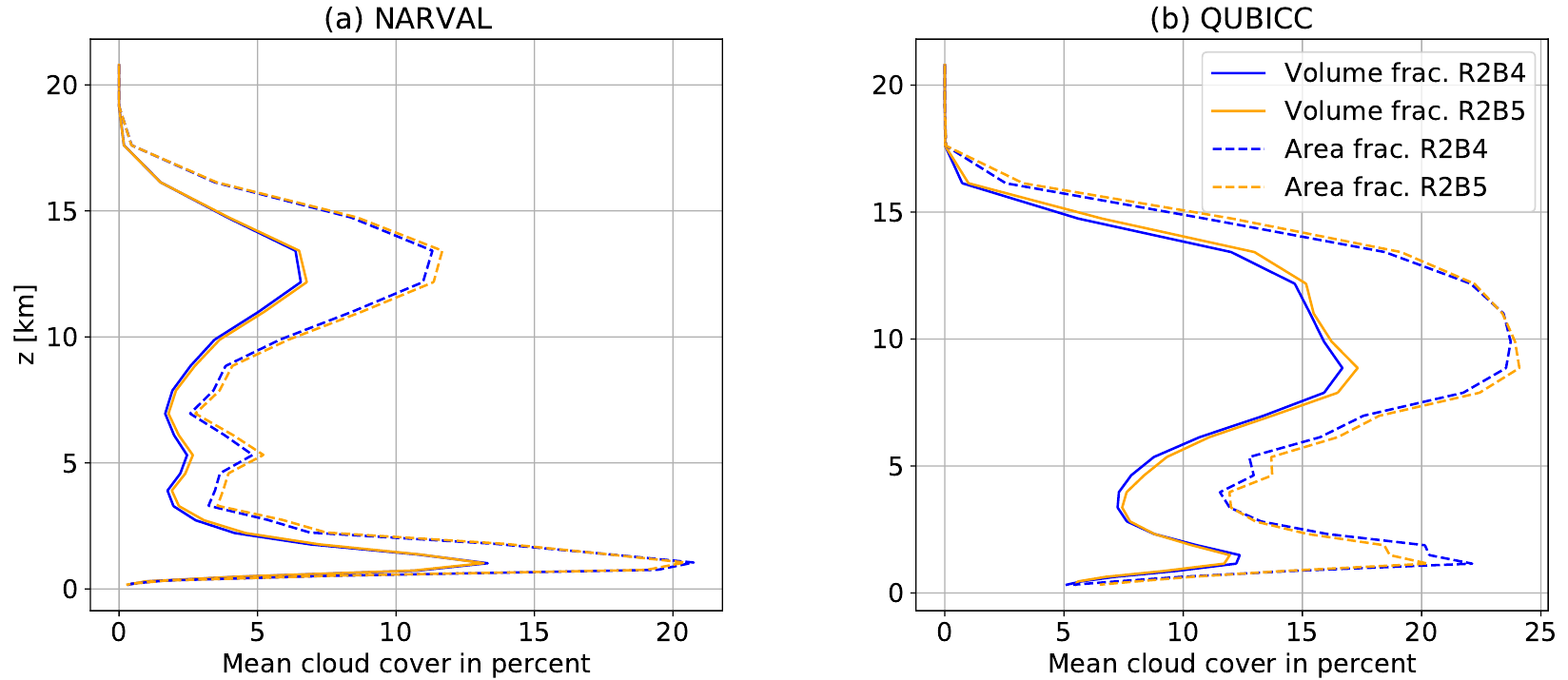}
\caption{Comparison of the coarse-grained mean cloud volume and mean cloud area fraction profiles for a) NARVAL and b) QUBICC. In a given grid cell, the cloud volume fraction is never greater than the cloud area fraction. Close to the surface, the grid cell thickness and thus also the vertical sub-grid variability of clouds is small. There it follows that the cloud area fraction is approximately equal to the cloud volume fraction.}
\label{fig:cloud_cover_vs_area}
\end{figure}

Equation (\ref{cloud_area_fraction}) is exemplified in panel b) of Figure \ref{fig:visualize_cg}. For QUBICC grid cells, which are always either fully cloudy or cloud-free, we can directly interpret equation (\ref{cloud_area_fraction}) as returning the fraction of high-resolution horizontal grid points that are covered by a cloud of any non-zero vertical extent within a coarse vertical cell. Due to the fractional cloudiness and the maximum overlap assumption, this link is less direct for the NARVAL data. Figure \ref{fig:cloud_cover_vs_area} illustrates the different mean vertical profiles of cloud volume fraction and cloud area fraction. Considerable differences in their coarse-grained vertical profiles (differing absolutely by almost $10\%$ on some layers) corroborate the need to distinguish these two concepts of cloud cover.

\section{Comparing Two Neural Networks Using Attribution Methods} \label{app:shap}
We use SHAP to compare two neural networks and to decompose model errors. However, our error decomposition framework can be used with any attribution method (LRP, LIME, integrated gradients, etc., \citeA{samek2019}) which fulfills the property that the attributed feature importances sum up to the predicted model output (possibly shifted by a constant value).

For a given NN $h$, data sample $X$ and input feature $i$, the SHAP package computes the corresponding Shapley value $\phi_{h, X, i}$. Shapley values satisfy the so-called efficiency property for every sample, which means that they sum up to the difference between the model output and its \textit{base value} (the expected model output)
\begin{equation} \label{a1}
\sum_{i \in I} \phi_{h, X, i} = h(X) - \mathbb{E}[h(X)],
\end{equation}
where $I \subseteq \mathbb{N}$ consists of the features' indices. A Shapley value $\phi_{f, X, i}$ can thus be interpreted as the amount by which an input feature $i$ contributes to the deviation of $f$'s prediction from the base value. Shapley values are constructed so that $f(X) - \mathbb{E}[f(X)]$ is fairly distributed among the features. \\
Let $f$ be the QUBICC R2B5 and $g$ the NARVAL R2B4 NN. Their base values $B_f := \mathbb{E}[f(X)]$ and $B_g := \mathbb{E}[g(X)]$ are computed as the average prediction of $f$ and $g$ on a subset of their respective training data sets (the so-called \textit{background data set}). By repeatedly drawing an appropriate sample from the training set of $f$, we can construct its background data set such that $B_f = B_g$. Plugging $f$ and $g$ into (\ref{a1}) we get
\begin{equation} \label{a2}
\sum_{i \in I} \phi_{f, X, i} - \sum_{j \in J} \phi_{g, X, j} = f(X) - g(X) + B_f - B_g = f(X) - g(X),
\end{equation}
where $I, J \subseteq \mathbb{N}$. Let $S$ be a random subset of the NARVAL R2B5 data and the overline $\overline{\cdot}$ denote the average over all samples in $S$. The size of $S$ is chosen to be large enough such that \textbf{i)} $\overline{f}$ and $\overline{g}$ are good approximations of the predicted averages of $f$ and $g$ on the entire NARVAL R2B5 data set (as shown in Figures \ref{fig:r2b5_qubicc_on_narval}a and S5a) and \textbf{ii)} the mean Shapley values are robustly estimated. \\ 
The sum of Shapley values corresponding to input features that are present in only one model (such as $\rho$) are in our case very small (absolute value $<0.08$) and thus negligible. Hence, by averaging over (\ref{a2}) we can approximate the mismatch between the average outputs of $f$ and $g$ by the sum of the difference of averaged Shapley values corresponding to features that $f$ and $g$ have in common
\begin{eqnarray} \label{shap_eqns}
\overline{f} - \overline{g} &=& \sum_{i \in I \cap J}(\overline{\phi_{f, X, i}} - \overline{\phi_{g, X, i}}) + \sum_{i \in I \setminus J} \overline{\phi_{f, X, i}} - \sum_{i \in J \setminus I} \overline{\phi_{g, X, i}} \\
&\approx& \sum_{i \in I \cap J}(\overline{\phi_{f, X, i}} - \overline{\phi_{g, X, i}}). \nonumber
\end{eqnarray}
So by comparing $\overline{\phi_{f, X, i}}$ and $\overline{\phi_{g, X, i}}$ for all common features $i \in I \cap J$ individually, we can explain which input features contribute to the difference between $\overline{f}$ and $\overline{g}$. Having ensured that $S$ satisfies \textbf{i)} and \textbf{ii)}, we can generalize (\ref{shap_eqns}) to the entire NARVAL R2B5 data set.

\acknowledgments
The neural network and analysis code can be found at \url{https://github.com/agrundner24/iconml_clc} and is preserved at \url{DOI: 10.5281/zenodo.5788873}. Primary data used in this work is archived by the Max Planck Institute for Meteorology (contact: \url{marco.giorgetta@mpimet.mpg.de}). The coarse-grained model output used for training the neural networks amounts to several TB. An extract from the training data is made available in the GitHub repository. The software code for the ICON model is available from \url{https://code.mpimet.mpg.de/projects/iconpublic}.

Funding for this study was provided by the European Research Council (ERC) Synergy Grant “Understanding and Modelling the Earth System with Machine Learning (USMILE)” under the Horizon 2020 research and innovation programme (Grant agreement No. 855187). Gentine acknowledges funding from LEAP NSF Science and Technology Center. We thank the Max Planck Institute for Meteorology for providing access to the NARVAL simulation data. Further we acknowledge PRACE for awarding us access to Piz Daint at ETH Zurich/CSCS, Switzerland, which made the QUBICC simulations possible (ID 2019215178). This work used resources of the Deutsches Klimarechenzentrum (DKRZ) granted by its Scientific Steering Committee (WLA) under project ID bd1083 and bd1179.


%
%

\bibliography{bibfile}

%
%
%
%
%

\end{document}


%
%


\graphicspath{{figures/}}

\title{Supporting Information for ``Deep learning based cloud cover parameterization for ICON"}
%
%

%
%


\authors{Arthur Grundner\affil{1,2}, Tom Beucler\affil{3}, Pierre Gentine\affil{2}, Fernando Iglesias-Suarez\affil{1}, Marco A. Giorgetta\affil{4}, and Veronika Eyring\affil{1,5}}

\affiliation{1}{Deutsches Zentrum für Luft- und Raumfahrt e.V. (DLR), Institut für Physik der Atmosphäre, Oberpfaffenhofen, Germany}
\affiliation{2}{Columbia University, Center for Learning the Earth with Artificial intelligence And Physics (LEAP), New York, NY 10027, USA}
\affiliation{3}{University of Lausanne, Institute of Earth Surface Dynamics, Lausanne, Switzerland}
\affiliation{4}{Max Planck Institute for Meteorology, Hamburg, Germany}
\affiliation{5}{University of Bremen, Institute of Environmental Physics (IUP), Bremen, Germany}



%
%

%

\begin{article}

\noindent\textbf{Contents}
\begin{enumerate}
\item Text S1 to S3
\item Figures S1 to S6
\item Tables S1 to S2
\end{enumerate}

\begin{figure}
\centering
\hspace*{-4em}\includegraphics[scale=0.43]{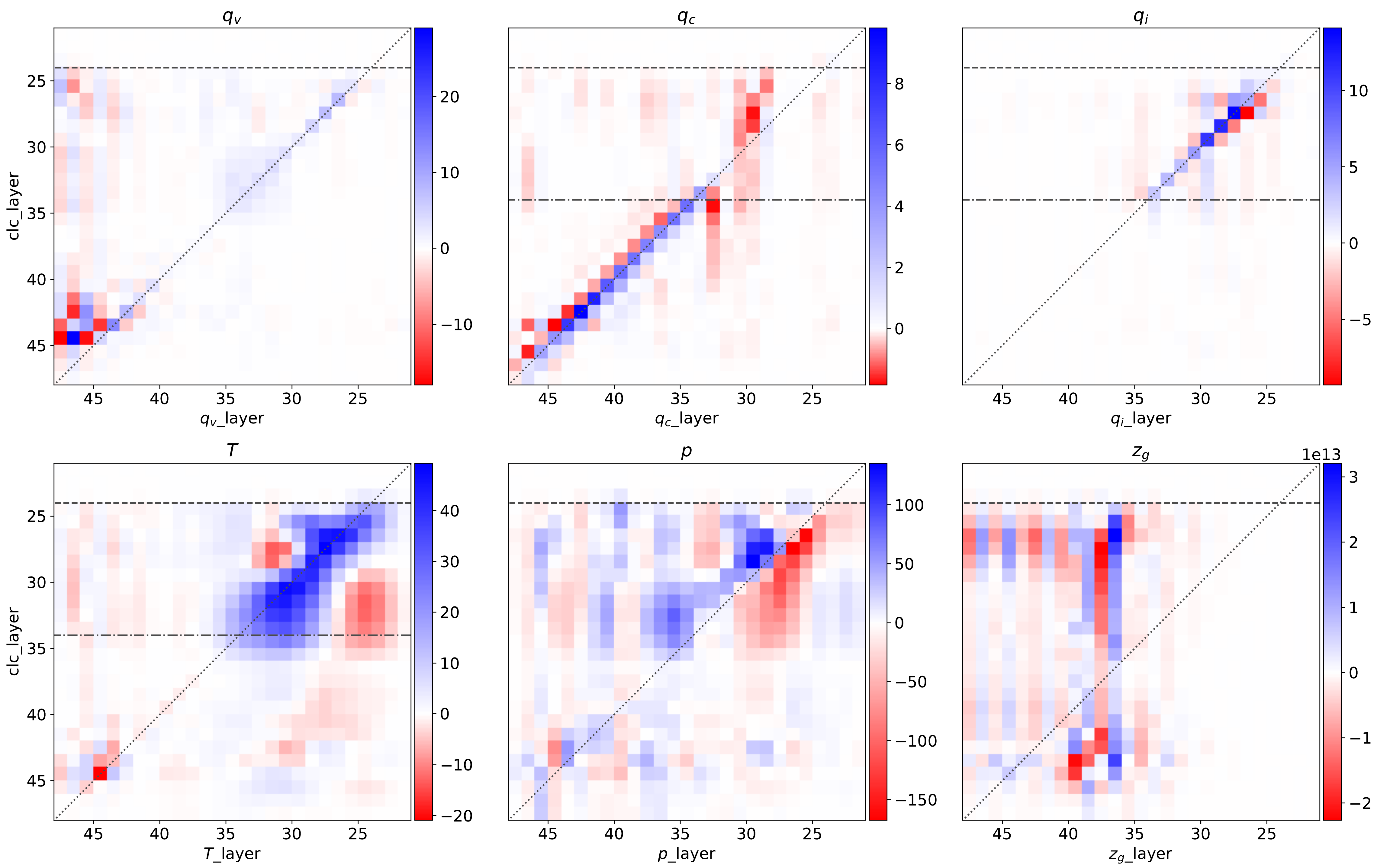}
\caption{Coefficients of the best multiple linear model on standardized NARVAL R2B4 data. The dashed line shows the tropopause ($\approx$\,$15$\,km), the dash dotted line shows the freezing level (i.e. where temperatures are on average below 0 degrees) ($\approx$\,$5$\,km) and the dotted line visualizes the diagonal. The coefficients suggest that the problem of diagnosing cloud cover is non-local. The zg coefficients seem to dominate. An elevated grid cell on level 15 increases cloud cover significantly. However, due to the nature of the vertical grid, the layers below will also be elevated, driving a decrease of cloud cover. An increase in specific humidity, cloud water (at altitudes below the freezing level) and cloud ice (at altitudes above the freezing level) increase cloudiness in the same grid cell. In the upper troposphere, when we increase the pressure, we force the condensation of water vapor at the given level and above.}
\label{fig:multilin_coefs}
\end{figure}
\begin{figure}
\centering
\includegraphics[scale=0.6]{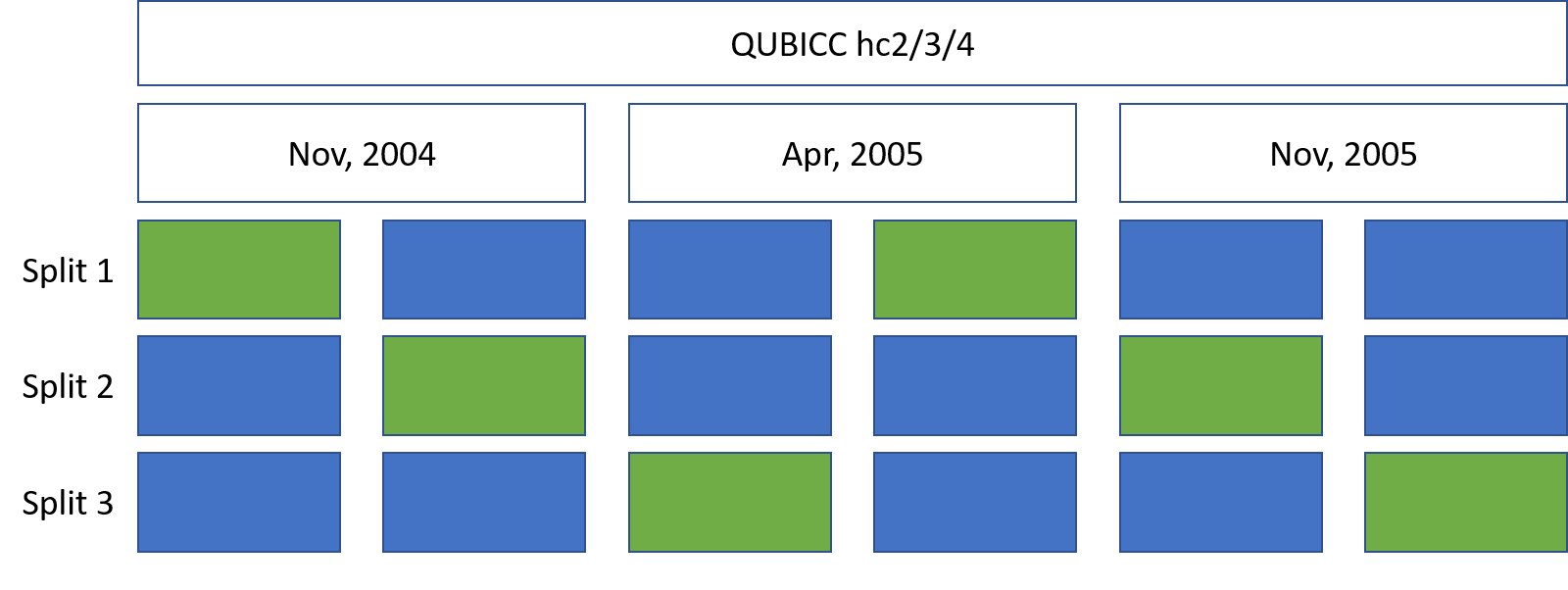}
\caption{We split the R2B5 data using a three-fold temporally coherent cross-validation split. In each split, we train a network on the blue folds and validate it on the green folds. One fold covers approximately 15 days.}
\label{fig:cross_validation_split}
\end{figure}
\begin{figure}
\centering
\includegraphics[scale=0.8]{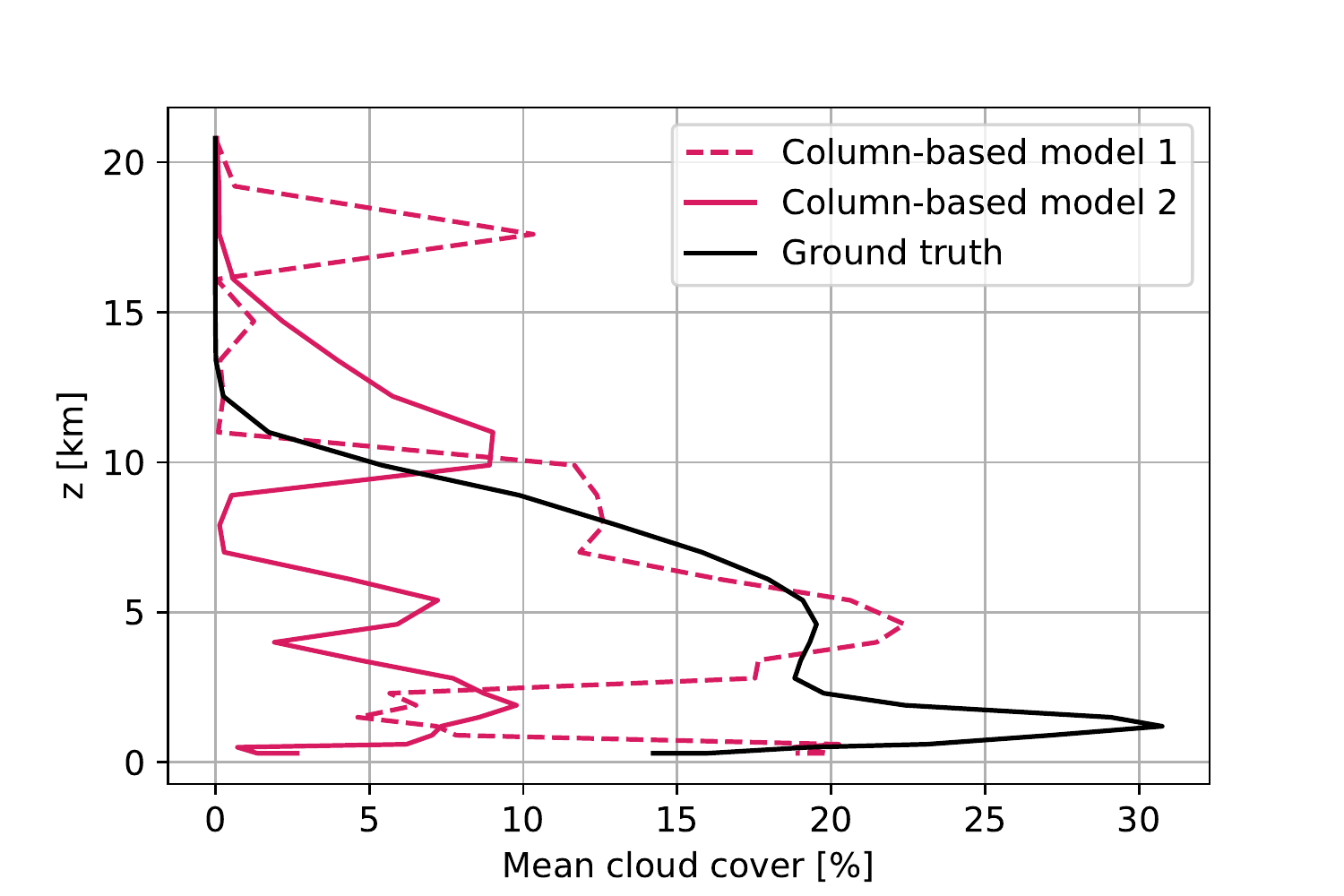}
\caption{Two different column-based models trained on NARVAL R2B4 data evaluated on QUBICC R2B4 data over the Southern Ocean and Antarctica ($< 60$S). Models from the same type stop being consistent and deviate significantly from the ground truth.}
\label{fig:r2b4_southern_ocean_transfer}
\end{figure}
\begin{figure}
\centering
\hspace*{-3.2em}\includegraphics[scale=.9]{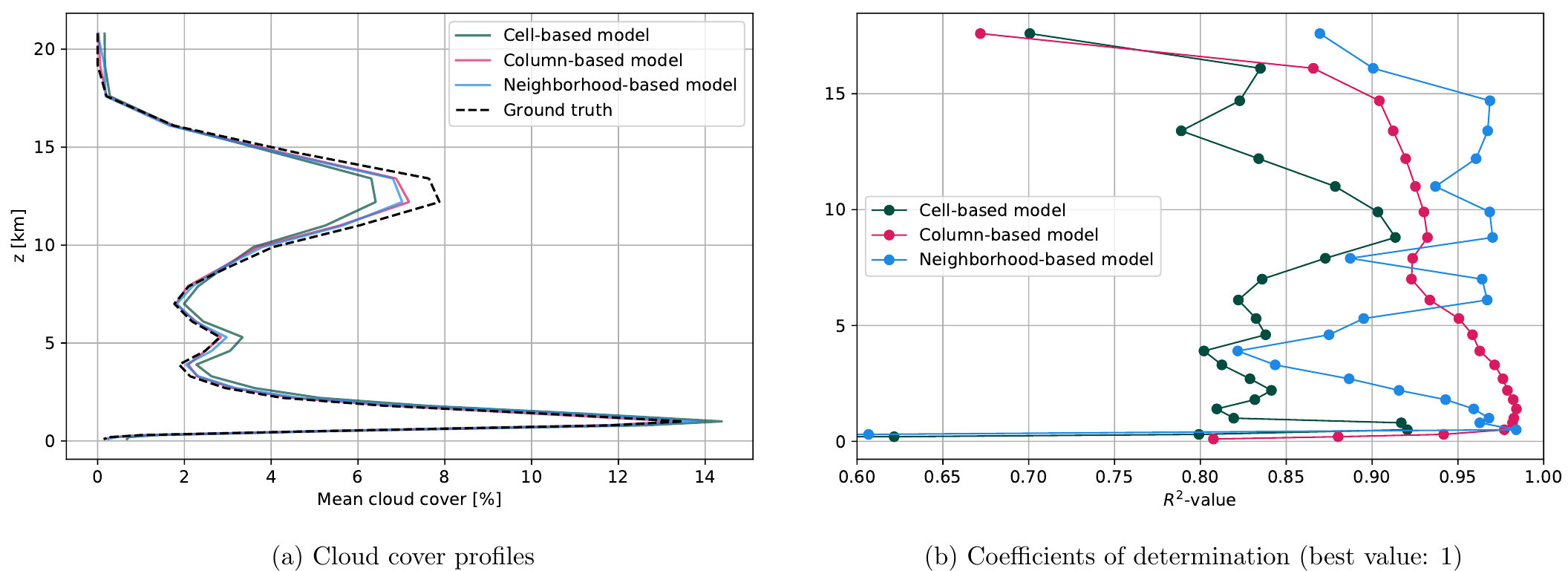}
\caption{The NNs trained on NARVAL R2B4 data evaluated on the coarse-grained and preprocessed NARVAL R2B5 data.}
\label{fig:r2b4_narval_on_r2b5}
\end{figure}
\begin{figure}
\centering
\hspace*{-5em}\includegraphics[scale=.3]{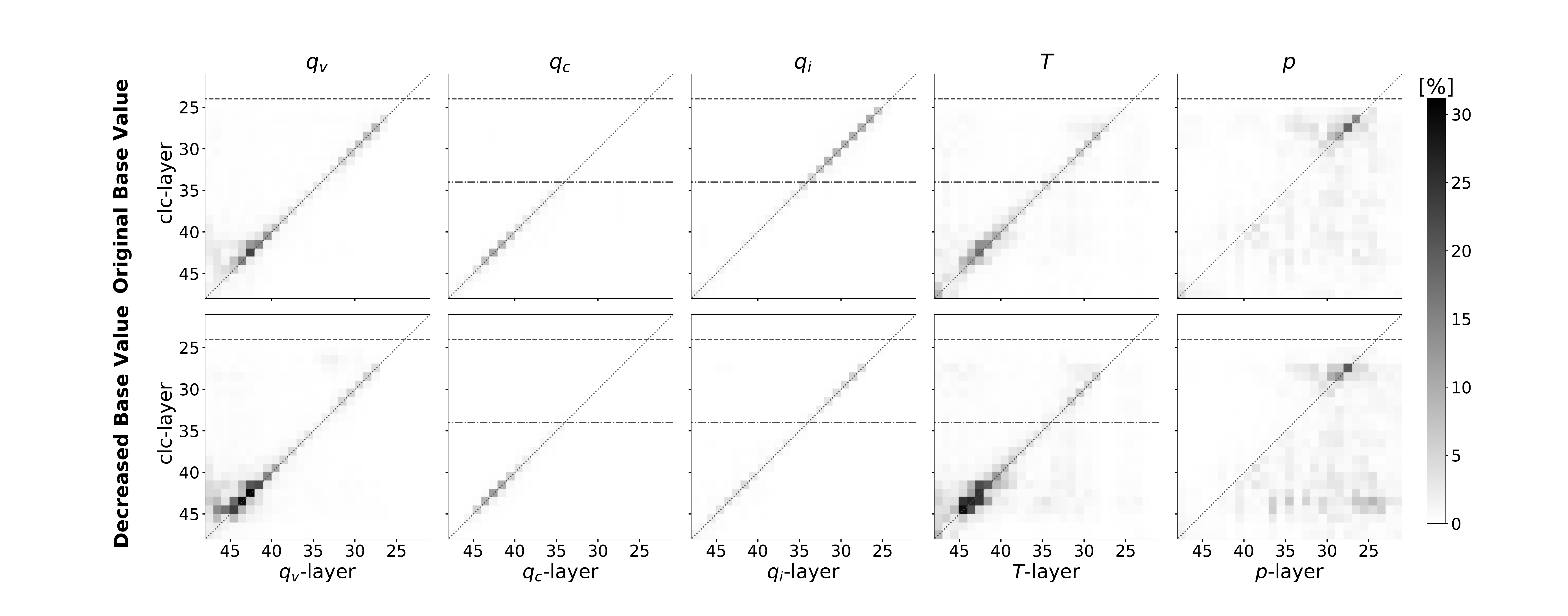}
\caption{Average absolute SHAP values of the QUBICC R2B5 column-based model when applied to a sufficiently large subset of the NARVAL R2B5 data. By repeatedly drawing an appropriate training sample from the QUBICC training data we decrease its base values, aligning them closely with the cloud cover profile of the NARVAL R2B5 data. Tests with ten different seeds have shown the values from the lower row to be robust, with pixel values not differing absolutely by more than $1$ or relatively by more than $20\%$. The input features that are not shown exhibit smaller absolute SHAP values ($z_g < 0.8\%$, \textit{land}/\textit{lake} $< 0.22\%$) everywhere and are thus omitted.}
\label{fig:average_absolute_SHAP_values_base_value_comp}
\end{figure}

 \noindent\textbf{Introduction}
This supplementary information provides more detailed information concerning the data and the neural networks (NNs). It describes the variables that were used as input features for the NNs, illustrates the architecture of the three NN types, the space of hyperparameter we explored and the preprocessing and amount of (training) data for each network. Table \ref{tab:param_packages} specifies the parameterization schemes used in the NARVAL and QUBICC simulations. The cross-validation split for the QUBICC (R2B5) models is depicted in Figure \ref{fig:cross_validation_split}. Figure \ref{fig:multilin_coefs} illustrates the coefficients of a multiple linear model trained on the NARVAL (R2B4) data. Figures \ref{fig:r2b4_southern_ocean_transfer} and \ref{fig:r2b4_narval_on_r2b5} cover aspects of the generalization capability of the NARVAL networks across regions and resolutions. Lastly, Figure \ref{fig:average_absolute_SHAP_values_base_value_comp} shows that SHAP values do not strongly depend on the base value.

------------------------------------------------------------------------ 

\section{Definition and Choice of Input Parameters for the NNs}
\begin{enumerate}
\item \textbf{land}: The land fraction (in $ [0, 1]$) is used in the ICON-A cloud cover scheme to discern whether one might have to artificially increase relative humidity in order to take thin maritime stratocumuli into account.
\item \textbf{lake}: The lake fraction (in $[0, 1]$) is a parameter closely related to the land fraction. A supply of moisture from the ground very likely influences the distribution of moisture in the atmospheric column above, especially in the presence of convection.
\item \textbf{Cor.}: The Coriolis parameter (in $1/s$) allows the cloud cover parameterization to vary between different latitudes, which can be especially useful with global training data.
\item $\mathbf{q_v}$, $\mathbf{T}$, $\mathbf{p}$, $\mathbf{z_g}$: Specific humidity (in $kg/kg$), air temperature (in $K$), pressure (in $Pa$) and geometric height at full levels (in $m$). These are the most important input variables for the original ICON-A cloud cover scheme (to compute relative humidity).
\item $\mathbf{q_c}$, $\mathbf{q_i}$: The specific cloud water content and the specific cloud ice content (in $kg/kg$). They have a direct influence on cloudiness as the presence of cloud water or ice is a necessary requirement for the presence of clouds. In this spirit, they are for instance used in an alternative 0-1 cloud cover scheme in ICON-A, which sets cloud cover to 1 when a certain threshold of cloud condensate is crossed.
\item {\boldmath$\rho$}: Air density (in $kg/m^3$). We left it out for the R2B5 NNs, since air density can mostly be derived from $p$, $T$ and $q_v$ by using the ideal gas law and is therefore redundant.
\item $\mathbf{u}$, $\mathbf{v}$: Zonal/eastward wind and meridional/northward wind (in $m/s$). Vertical wind shear can induce a large difference between cloud area fraction and cloud cover.
\item $\mathbf{clc_{t-1}}$: The cloud cover estimate (in $[0, 100]$\%) from the previous timestep (1 hour before). Undeniably, clouds have a memory effect on this time scale. However, a model that relies on previous cloudiness cannot be used in the first time step.
\end{enumerate}
The features $\rho$, $u$, $v$ are also used in the Tompkins scheme of cloud cover \cite{tompkins2002}.

\section{Preprocessing}
The preprocessing, which we define as distinct from coarse-graining, consists of up to four steps:
\begin{enumerate}
\item \textbf{For all cell-based and QUBICC neighborhood-based models (N1, Q1 and Q3)}: Ensure that the amount of data samples with $clc \neq 0$ is as large (for the Q1 model twice as large to reduce the data size) as the one with $clc = 0$, by downsampling the latter class of cloud-free data samples. 
\item \textbf{For the neighborhood-based NARVAL models (N3)}: Remove the cloud cover from the first time step of each day of the NARVAL data from the output. We cannot predict it, because there is no previous cloud cover value which the neighborhood-based NARVAL model would require as input.
\item \textbf{QUBICC data}: Remove the first time steps of the simulations because that output incorrectly consists of an entirely cloud-free atmosphere. Scale the cloud cover to be in $[0, 100]\%$. Convert the data from float64 to float32 to reduce the data size.
\item \textbf{For the QUBICC cell- and neighborhood-based models (Q1 and Q3)}: Subsample only every third hour from the QUBICC data set to reduce the data size. Assuming a high temporal correlation, we should not lose a lot of information. Remove condensate-free clouds ($\sim 7\%$ of all clouds).
\item \textbf{For all models (N1-N3, Q1-Q3)}: Normalize the actual training data so that each input feature to the NN is distributed according to a Gaussian with zero mean and unit variance. In the column-based models this means that the normalization is done on a level-by-level basis and for the cell-based and neighborhood-based models we have one level-independent mean and standard deviation per input feature. According to \citeA{brenowitz2019}, we expect the impact on our results due to these different choices of normalization to be very small. This step of normalization can only be done after splitting the set of all training data samples into subsets of training, validation and test data.
\end{enumerate}

\section{Space of Hyperparameters}
We explored the following space of hyperparameters used in the neural network training:
\begin{enumerate}
\item Number of units per hidden layer: 16, 32, ..., 512 
\item Number of hidden layers: From 1 to 4
\item Activation functions: ReLU, ELU, tanh, leaky ReLU with $\alpha \in \{0.01, 0.2\}$
\item Initial learning rate: From $10^{-4}$ to 1
\item Epsilon parameter in the optimizer: $10^{-8}$, $10^{-7}$, $0.1$, 1
\item Dropout: With or without after each hidden layer with parameters from 0 to 0.5
\item L1/L2-regularization: With parameters from 0 to 0.01
\item Batch normalization: With or without after each layer
\end{enumerate}

\bibliography{bibfile}

------------------------------------------------------------------------ 


\end{article}
\clearpage


%
%
%
%
%
%
%
%
%

\begin{figure}
    \centering
    \hspace*{-0.5cm}\includegraphics[scale=0.54]{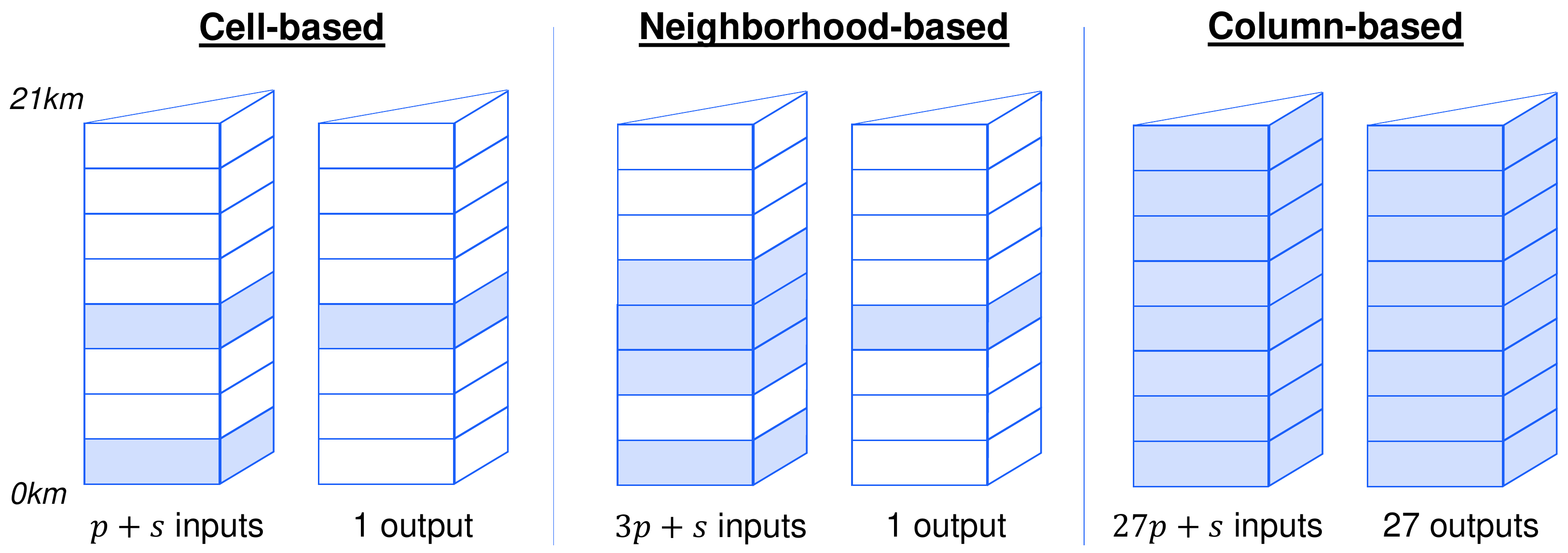}
    \caption{A sketch of the three NN types based on one grid column. The variable $p$ denotes the number of input features from the grid cells and $s$ is the number of extra variables from the surface. In this sketch, the neighborhood-based model uses two neighboring cells, which is only true for our QUBICC-trained NN.}
    \label{fig:three_nn_types}
\end{figure}

\hspace{1em} \\
\hspace{1em} \\

%
%
%


\begin{table}
\centering
\caption{Parameterizations used in the NARVAL and QUBICC simulations}
\label{tab:param_packages}
\rowcolors{2}{}{gray!10}
\begin{tabular}{ l c c }
 & \textbf{NARVAL} & \textbf{QUBICC} \\
 \midrule
 \textbf{Cloud Cover} & Diagnostic PDF & \begin{tabular}{@{}c@{}}  All-or-nothing scheme \\ based on cloud condensate \end{tabular} \\
 \textbf{Microphysics} & \begin{tabular}{@{}c@{}} Single-moment scheme \\ \cite{doms2011description, seifert2008revised} \end{tabular} & \begin{tabular}{@{}c@{}} Single-moment scheme \\ \cite{doms2011description, seifert2008revised} \end{tabular} \\
 \textbf{Radiation} & \begin{tabular}{@{}c@{}} RRTM scheme \\ \cite{barker2003assessing, mlawer1997radiative} \end{tabular} & \begin{tabular}{@{}c@{}} RTE+RRTMGP scheme \\ \cite{pincus2019} \end{tabular} \\
 \textbf{Turbulence} & \begin{tabular}{@{}c@{}} Prognostic TKE \\ \cite{raschendorfer2001new} \end{tabular} & \begin{tabular}{@{}c@{}} Total turbulent energy scheme \\ \cite{mauritsen2007total} \end{tabular} \\
 \textbf{Land} & \begin{tabular}{@{}c@{}} Tiled TERRA \\ \cite{schrodin2001multi, schulz2015evaluation} \end{tabular} & \begin{tabular}{@{}c@{}} JSBach4-lite \cite{raddatz2007will} \end{tabular} \\
\end{tabular}
\end{table}




\begin{table}
\centering
\caption{Amount of training data samples for the NNs. The tuples denote either (time steps, vertical layers, horizontal fields) or (time steps, horizontal fields). Note that for the R2B4 neighborhood-based model we trained one NN per vertical layer, so the number of training samples is equal to the number of training samples for the R2B4 column-based model. Grid columns containing grid cells that were omitted during coarse-graining are excluded in the `After coarse-graining'-column and are also not used for training.}
\label{tab:training_data}
\rowcolors{2}{}{gray!10}
\begin{tabular}{ l l l c }
 \toprule
 & Original data ($\leq 21$\,km) & After coarse-graining & After preprocessing \\
 \midrule
  \textit{Cell-based} & & & \\
  R2B4 NARVAL & $5.6 \cdot 10^{11}$ $(1721, 66, 4887488)$ & $4.5 \cdot 10^7$ $(1635, 27, 1024)$ & $3.7 \cdot 10^7$  \\
 R2B5 QUBICC & $3.9 \cdot 10^{12}$ $(2162, 87, 20971520)$ & $4.6 \cdot 10^{9}$ $(2162, 27, 78069)$ & $8.8 \cdot 10^8$ \\
  \addlinespace
  \textit{Neighborhood-based} & & & \\
  R2B4 NARVAL & $8.4 \cdot 10^9$ $(1721, 4887488)$  & $1.7 \cdot 10^6$ $(1632, 1024)$ & $1.7 \cdot 10^6$ \\
 R2B5 QUBICC & $3.9 \cdot 10^{12}$ $(2162, 87, 20971520)$ & $4.6 \cdot 10^9$ $(2162, 27, 78069)$ & $1.2 \cdot 10^9$ \\
  \addlinespace
   \textit{Column-based} & & & \\
  R2B4 NARVAL & $8.4 \cdot 10^9$ $(1721, 4887488)$  & $1.7 \cdot 10^6$ $(1635, 1024)$ & $1.7 \cdot 10^6$  \\
 R2B5 QUBICC & $4.5 \cdot 10^{10}$ $(2162, 20971520)$ & $1.7 \cdot 10^8$ $(2162, 78069)$ & $1.7 \cdot 10^8$ \\
  \bottomrule
\end{tabular}
\end{table}